\documentclass[letterpaper]{article} % DO NOT CHANGE THIS
\usepackage{aaai24}  % DO NOT CHANGE THIS
\usepackage{times}  % DO NOT CHANGE THIS
\usepackage{helvet}  % DO NOT CHANGE THIS
\usepackage{courier}  % DO NOT CHANGE THIS
\usepackage[hyphens]{url}  % DO NOT CHANGE THIS
\usepackage{graphicx} % DO NOT CHANGE THIS
\usepackage{natbib}  % DO NOT CHANGE THIS AND DO NOT ADD ANY OPTIONS TO IT
\usepackage{caption} % DO NOT CHANGE THIS AND DO NOT ADD ANY OPTIONS TO IT
\usepackage{cite}
\usepackage{amsmath,amssymb,amsfonts}
\usepackage{textcomp}
\usepackage{xcolor}
\def\BibTeX{{\rm B\kern-.05em{\sc i\kern-.025em b}\kern-.08em
    T\kern-.1667em\lower.7ex\hbox{E}\kern-.125emX}}

\usepackage[caption=false]{subfig}
\usepackage{svg}
\usepackage{multicol}
\usepackage{multirow}
\usepackage{color}
\usepackage{hhline}
\usepackage{pifont}
\usepackage{comment}
\usepackage{cleveref}
\usepackage{tablefootnote}
\usepackage{booktabs}
\usepackage[ruled,vlined]{algorithm2e}
\definecolor{mynicegreen}{RGB}{11,102,35}
\usepackage[group-separator={,}]{siunitx}

\usepackage{amsthm}
\theoremstyle{definition}
\newtheorem{definition}{Definition}

\newcommand{\answerYes}[1]{\textcolor{blue}{#1}} 
\newcommand{\answerNo}[1]{\textcolor{teal}{#1}} 
\newcommand{\answerNA}[1]{\textcolor{gray}{#1}}

 \newcommand{\squishlist}{
	\begin{list}{$\bullet$}
		{ \setlength{\itemsep}{0pt}
			\setlength{\parsep}{3pt}
			\setlength{\topsep}{3pt}
			\setlength{\partopsep}{0pt}
			\setlength{\leftmargin}{1.5em}
			\setlength{\labelwidth}{1em}
			\setlength{\labelsep}{0.5em} } }
	
	\newcommand{\squishlisttwo}{
		\begin{list}{$\bullet$}
			{ \setlength{\itemsep}{0pt}
				\setlength{\parsep}{0pt}
				\setlength{\topsep}{0pt}
				\setlength{\partopsep}{0pt}
				\setlength{\leftmargin}{2em}
				\setlength{\labelwidth}{1.5em}
				\setlength{\labelsep}{0.5em} } }
		
		\newcommand{\squishend}{
	\end{list}}
\newcommand{\cmmnt}[1]{\ignorespaces}

\title{Wildlife Product Trading in Online Social Networks:\\A Case Study on Ivory-Related Product Sales Promotion Posts}
\author{
Guanyi Mou\equalcontrib,
Yun Yue\equalcontrib,
Kyumin Lee,
Ziming Zhang
}
\affiliations{
Worcester Polytechnic Institute\\
100 Institute Road, Worcester, MA, 01609\\
\{gmou, yyue, kmlee, zzhang15\}@wpi.edu
}

\begin{document}

\maketitle

\begin{abstract}

Wildlife trafficking (WLT) has evolved into a pressing global concern, as traffickers increasingly utilize online platforms such as e-commerce websites and social networks to expand their illicit trade. This paper addresses the pivotal challenge of detecting and recognizing promotional behaviors related to the sale of wildlife products within online social networks—a critical step in combating these environmentally detrimental activities. To confront these illicit operations effectively, our research undertakes the following key initiatives: 1. Data Collection and Labeling: We employ a network-based approach to gather a scalable dataset pertaining to wildlife product trading. Through a human-in-the-loop machine learning process, this dataset is meticulously labeled, distinguishing between positive class samples containing wildlife product selling posts and hard-negatives representing regular posts misclassified as potential WLT posts, subsequently rectified by human annotators. 2. Machine Learning Framework Development: We present a robust framework that benchmarks machine learning results on the collected dataset. This framework autonomously identifies suspicious wildlife selling posts and accounts, effectively harnessing the multi-modal nature of online social networks. 3. In-depth Analysis of Trading Behaviors: Our research delves into a comprehensive analysis of trading posts, illuminating the systematic and organized selling behaviors prevalent in the current landscape. By providing detailed insights into the nature of these behaviors, we contribute valuable information for understanding and countering illegal wildlife product trading. Moreover, we emphasize our commitment to openness and collaboration by making our code and dataset openly available, thereby fostering cooperative efforts towards the development of more effective strategies in combating illegal wildlife trafficking.
\end{abstract}

\section{Introduction}
\label{sec:intro}

Wildlife trafficking, defined as ``the poaching or other taking of protected or managed species and the illegal trade in wildlife and their related parts and products'',\footnote{\url{https://www.fws.gov/international/wildlife-trafficking/}} has evolved into a critical international crisis. Despite dedicated efforts from officials,\footnote{\url{https://www.traffic.org/about-us/legal-wildlife-trade/}} non-profit organizations (NGOs),\footnote{\url{https://www.eagle-enforcement.org/}} and researchers, the illicit trade of endangered wildlife persists globally, thriving in lucrative black markets~\cite{zimmerman2003black, moyle2009black, alacs2008wildlife}. 

\captionsetup[subfigure]{labelformat=empty}
\begin{figure}[tbp]
    \centering
    \small
    \subfloat[``A superb 18th century European carved ivory dish. $11\frac{1}{2}$in wide ... Estimate: £1500-2000 {{MENTION}} {{URL}}.'']{
        \includegraphics[width=.82\columnwidth]{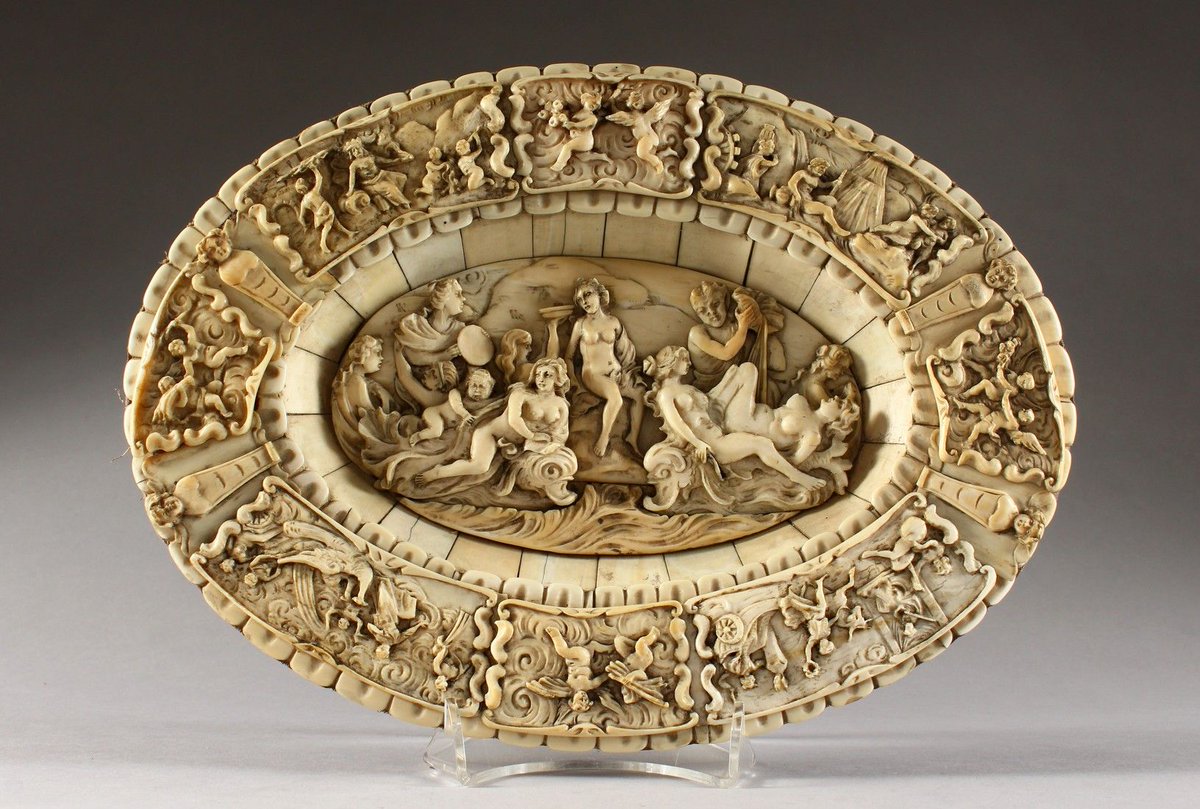}
    }
    \caption{An example for wildlife product trading related post in the Online social networks. The subcaption is the text content where links and user mentions are masked. We investigate whether a post is WLT-related through its post text, images, and the linked webpages whenever necessary. }
    \label{fig:wivory}
\end{figure}

\captionsetup[subfigure]{labelsep=period, labelformat=simple}

The chain of wildlife trafficking involves illegal capturing, organized transportation, and trading of wildlife and their products. This research concentrates on the final stage, wildlife trading, guided by the principle ``No trading, No killing''.\footnote{\url{https://tinyurl.com/yc786mh6}} While the preceding stages primarily operate offline, wildlife trading has recently intertwined with the internet~\cite{lavorgna2014wildlife}. The surge in online trading markets, especially through e-commerce websites, has exposed illegal wildlife trading to a wider audience worldwide~\cite{sung2018assessing}. Previous research has predominantly analyzed established online marketplaces like Etsy and Ebay~\cite{sinovas2017wildlife, pascual2021assessing, miller2019detecting}. In contrast, limited literature has explored the impact of wildlife trafficking on online social networks~\cite{xu2020illegal,xu2019use}. Moreover, we found the existing works analyzing wildlife trafficking in online social networks are preliminary, with a lack of exploration into the abundant hidden information, including the multi-modal nature of selling posts. Systematically studying and leveraging these aspects to combat online crimes remain challenging. The pieces of evidence are also limited due to the scarce distribution nature of wildlife product sales posts. For example, \citet{xu2019use} filtered 138,357 suspicious tweets on Twitter, only finding 53 tweets from 38 unique users involved in ivory selling and zero pangolin-related posts. The limited number of accounts/tweets also makes building real-life applicable, effective machine learning less convincing. 
Under such a low recall rate, manually identifying wildlife product sales posts can be a Sisyphean effort.
We urgently need a more efficient method to collect scalable data. The training and deployment of automatic learning algorithms that recognize potential illegal wildlife product trades can only become viable with such a method as a premise. Fig.~\ref{fig:wivory} illustrates this challenge, with masked image and text content,\footnote{There are some examples in this paper for illustrative purposes in research. Authors did not intend to advertise the sales of wildlife products and tried their best to protect privacy.} emphasizing the need for more efficient data collection methods.

To address these gaps, this research delves deeper into the patterns hidden behind selling posts on online social networks. For the first time, we present an Ivory-related multi-modal dataset, featuring positive samples and hard-negatives.\footnote{In this paper, we use ``positive class'' and ``WLT class'' interchangeably. Like wise, we use ``hard-negatives'', ``negative class'', and ``normal class'' interchangeably.} Leveraging a network-propagation-based method for data collection and a human-in-the-loop approach for labeling, our scalable and adaptive strategy can extend to other wildlife product categories with minimal effort. We benchmark machine learning results, introducing a practical framework that capitalizes on the multi-modality of the data. Furthermore, we offer rich observations and insights into distinguishing between the two classes, contributing to a more effective approach in identifying potential illegal wildlife product trades.

\smallskip\noindent To this end, we make the following contributions:
\squishlist
\item Data Scale Expansion: Beginning with minimal seed posts, we implement a network-propagation method to significantly expand the scale of our dataset, automating the collection of suspicious wildlife product trading (WLT) posts. This method is not only scalable but also highly adaptive, extending its applicability to various categories of WLT posts. 
\item Efficient Human-in-the-Loop Mechanism: We employ a human-in-the-loop mechanism to streamline the extraction of the most suspicious posts and their labeling. This approach efficiently reduces the laborious human effort required for identifying WLT posts and provides crucial hard negatives—normal/non-WLT posts that might be misclassified as WLT posts.
\item First Scalable Dataset for OSNs\footnote{\url{https://github.com/GMouYes/WLT-OSN}}: We introduce and share a dataset focused on wildlife trading-related posts within online social networks (OSNs). To the best of our knowledge, this dataset is the first of its kind, offering scalability and empowering machine learning algorithms to automatically identify potential WLT posts—scarce yet profoundly impactful to the global ecosystem.
\item Benchmarking Automatic Machine Learning Results: Using our dataset, we conduct a comprehensive benchmarking of automatic machine learning results, considering various modalities and design options. Our paper presents state-of-the-art results, supported by multiple evaluation metrics, advancing the field's understanding of WLT detection.
\item Systematic Analysis of Patterns: We systematically analyze the distinctive patterns of WLT posts in comparison to hard-negative normal posts within online social networks. Our analysis provides valuable insights from multiple perspectives, serving as a foundation to attract the interest of researchers and paving the way for future works in this critical domain.
\squishend

\section{Related Work}
\label{sec:related}

Wildlife trafficking, acknowledged as the ``Second-biggest direct threat to species after habitat destruction'' by the World Wildlife Fund (WWF),\footnote{\url{https://tinyurl.com/2p89zahk}} poses a severe global challenge. Over the years, illegal activities involving the capture and sale of endangered species have been meticulously documented. \citet{wyatt2021wildlife} extensively detailed the organized crime dynamics, profiling both offenders and victims. Legislators have also delved into the complexities of wildlife trafficking situations~\cite{sollund2019crimes}. Esteemed organizations like WWF,\footnote{\url{https://www.worldwildlife.org/}} TRAFFIC,\footnote{\url{https://www.traffic.org/}} and EAGLE\footnote{\url{https://www.eagle-enforcement.org/}} actively combat violators engaged in the systematic illegal hunting, transportation, and sale of wildlife and their products~\cite{dalberg2012fighting}.

Historically, significant efforts have been invested in providing data~\cite{gore2022voluntary, gore2023data} and conducting analyses~\cite{gore2019transnational, gore2023advancing} focusing on offline trafficking. However, with the advent of the internet, wildlife traffickers have adapted their strategies, actively promoting and selling products online~\cite{lavorgna2014wildlife}. Most existing research and organizational reports concentrate on analyzing large online marketplaces~\cite{sinovas2017wildlife, pascual2021assessing, miller2019detecting, alfino2020code,cardoso2023detecting}. For instance, \citet{miller2019detecting} provide a systematic report on wildlife product trafficking across multiple online platforms. \citet{alfino2020code} present intriguing observations, noting that traffickers across European countries employ similar ``code words'' to facilitate cross-national trading.

While some studies explored wildlife product trading on the dark web, uncovering minimal evidence~\cite{harrison2016assessing,roberts2017bycatch}, only a handful have investigated promotions and sales on social networks (e.g., Twitter, Facebook, Instagram, Pinterest). \citet{wyatt2022wildlife} reported preliminary analysis on 500 private messages on Facebook and WhatsApp marketing `exotic' pets collected from RENCTAS.\footnote{\url{https://renctas.org.br/home-en/}} However no data is publicly available and data collection methodologies remain unknown.
\citet{xu2019use, xu2020illegal} conducted preliminary research on Facebook and Twitter, identifying a mere 53 tweets promoting ivory-related products out of 138,357 collected. The low evidence recall rate (0.038\%) highlights the inefficiency of manual verification in addressing wildlife promotion behaviors. Efficient methods surpassing simple keyword filtering mechanisms are urgently needed.
% \citet{di2019framework} proposed a general idea of building a possible framework for identifying wildlife promotion posts. Still, they did not provide or collect data, build specific models, or run experiments in the mentioned paper. 

Our work differs from the prior work:
1) We devise a novel method for efficiently collecting and labeling a scalable dataset. The approach can be applied to other works with minimum adaptation effort.
2) We form and share the first dataset for WLT posts in OSNs, enabling possible automatic learning algorithms to be trained upon and applied to in-the-wild WLT posts.
3) We systematically analyze the unique characteristics of WLT posts in OSNs. We point out the unique multi-modality nature of the problem. Finally, we benchmark the dataset with multiple baselines and provide the current state-of-the-art design, which effectively identifies scarce however damaging, WLT posts.

\section{Problem Formulation}
\label{sec:definition}
\begin{definition}
    \textbf{WLT post:} We define a wildlife product trading (WLT) related post on an online social network as a post that contains two key components: 
\squishlist
\item Discussion around wildlife products;
\item Discussion around selling/buying these products.
\squishend
\end{definition}
\noindent For example, Fig.~\ref{fig:wivory} explicitly shows a product made from ivory that is available for sale. On the other hand, only mentioning ivory (e.g., for education purposes) or selling other non-WLT-related products (e.g., wood) would be a normal post. We show more examples and discussions around the typical positive and negative posts in Sec.~\ref{sec:case_study}. In this work, we focus on ivory-related products and leave other types of wildlife products for future research.
\begin{definition}
    \textbf{WLT post identification:} Given OSN posts $X$ and label set $Y = \{0,1\}$ with the following property
    \begin{equation}
        x = \{T, I, A\}, x \in X
        \label{eq:def_post}
    \end{equation}
    where $x$ contains the text $T$, images $I$, and other attributes $A$, such as user descriptions or other behavioral features. 
    $X$ is naturally multi-modal and multi-dimensional. The label with value 1 (positive class) represents WLT-related posts, while value 0 (negative class) represents normal posts.
    The WLT-related post-detection task aims to find the optimal model $f_{\theta}:x \rightarrow y$, which maps $X$ to $Y$, and $\theta$ is the learnable parameters of the model.
\end{definition}

\section{Methodology}
\label{sec:methodology}
This section describes how we automatically collected suspicious data and adopted a human-in-the-loop method for labeling WLT posts and their hard-negative counterparts. We note the following techniques are scalable (i.e., further expandable given the current data), adaptable (i.e., applicable to other wildlife products), and efficient compared to keyword filtering or naive human labeling methods.

\begin{figure}[tbp]
    \centering
    \includegraphics[width=.85\linewidth]{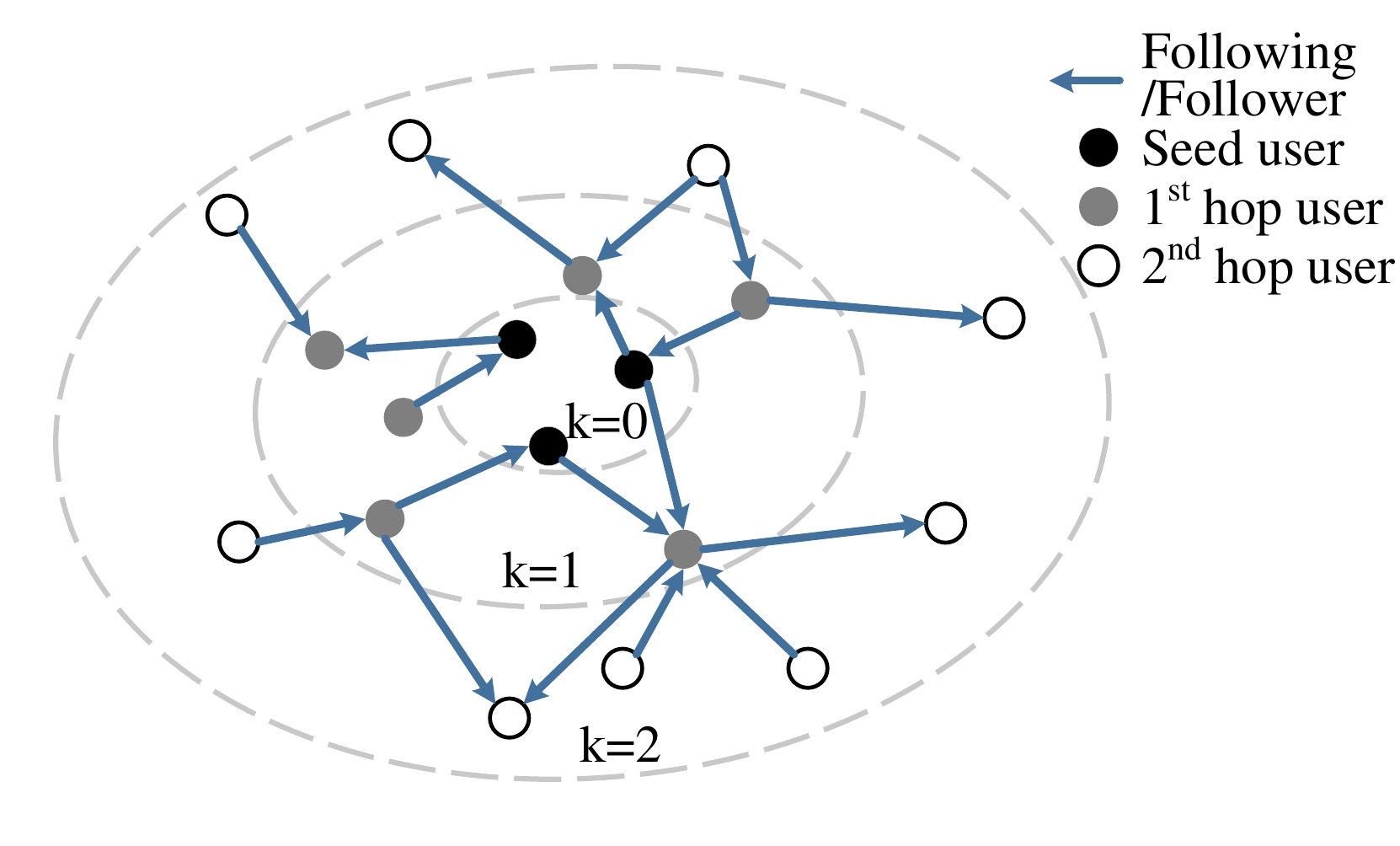}
    \caption{Illustration for Collecting Data. Nodes are the users, and edges represent their relationships. Given seed posts as nodes in black, we fetch their following/follower (blue and yellow edges) network for several hops. Eventually, we collect all these users' timelines as candidate data for further processing. Ideally, researchers can keep expanding the dataset scale by extracting more user hops, given their budget and computation limits.}
    \label{fig:data_collect}
\end{figure}

\subsection{Automate Data Collection}
\label{sec:data_collection}

We present a comprehensive overview of our data collection process in Fig.~\ref{fig:data_collect}. Commencing with nine seed tweets shared by \citet{xu2019use} pertaining to ivory-related Wildlife Product Trading (WLT) posts on Twitter, we acknowledge the significance of expanding the scale beyond these limited positive-only instances for the development of a practical and well-generalized machine learning application.

To address this limitation, we employ a network propagation approach to collect additional suspicious posts. Starting with the seed tweets, we retrieve the posting users, referred to as ``seed users'' and expand the network by collecting their followings, followers, and subsequent network layers. This scalable expansion process continues until the desired number of users is reached. Subsequently, we gather historical posts (up to 3,200 per user) from the collected users, resulting in a substantial dataset of suspicious posts for further selection and labeling.

\smallskip\noindent The collection process is guided by two key insights derived from preliminary analysis:
\squishlist
\item A user making one WLT post is likely to produce more WLT posts, justifying the search for additional posts from the same user.
\item WLT sellers are more likely to be connected within the network, as individuals or groups of sellers may control multiple accounts involved in the same business.
\squishend
Practically, we fetch the seed users and their two-hop neighbors based on the follower/following relationship and retrieve their posts, thus yielding an extensive dataset, comprising over \underline{\textit{ten million}} posts awaiting further processing.

\begin{figure}[tbp]
    \centering
    \includegraphics[width=.8\linewidth]{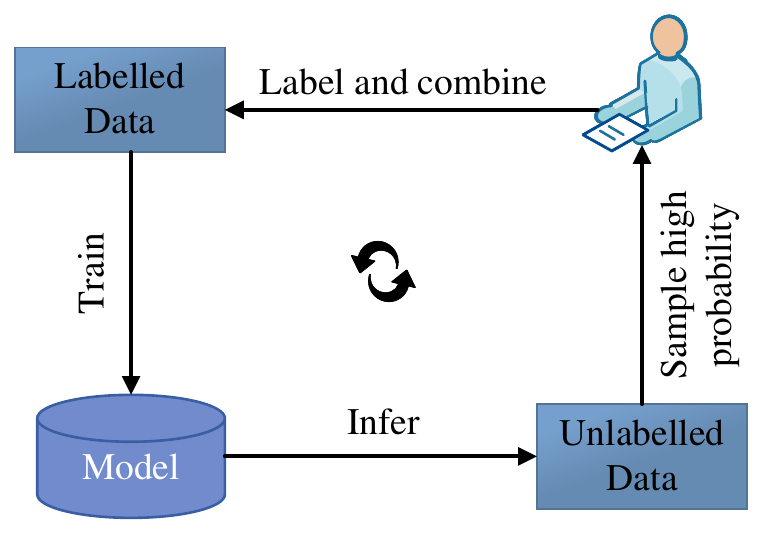}
    \caption{A human-in-the-loop process for labeling data.}
    \label{fig:labelling}
\end{figure}

\subsection{Human-in-the-loop Selection and Labeling}
\label{sec:data_label}

Although a substantial amount of data related to WLT is collected, the challenge lies in the large dataset size and a still-scarce positive rate, presenting an obstacle for human annotators. To overcome this, we implement an effective and efficient method for sampling and selecting highly probable WLT posts and labeling them.

\subsubsection{Algorithm Design and Labeling}
The human-in-the-loop data selection and labeling mechanism, depicted in Fig.~\ref{fig:labelling}, involves the following steps:
\begin{enumerate}
\item Initial sampling of the most recent $N$ posts from seed users, manually labeled, and combined with the seed posts to form the first labeled data group ($N=100$).
\item Division of the labeled data into train/test sets for machine-learning model training.
\item Model inference on a large proportion of unlabeled data, assigning probability scores between 0 and 1 for each post's likelihood of being WLT.
\item Selection of the top $K$ highest probability unlabeled data for another round of human labeling ($K \approx 2,500$).
\item Iterative repetition of steps 2-4 until the desired number of labeled posts ($\# posts \geq n$, where $n=8K$).
\item Combining all labeled data, cleaning and filtering (with a focus on English-based content) the dataset.
\end{enumerate}
During the labeling process, two of our authors served as annotators, bringing domain expertise and actively participating in discussions to define labeling criteria. Annotators independently labeled posts, with mutually agreed-upon labels being considered for adoption. Conflicting annotations were excluded from the final dataset.
It's noteworthy that the human-in-the-loop data selection and labeling process is scalable. As more rounds are executed, machine learning models capture better signals for recognizing WLT posts, and human annotators label more highly suspicious posts. In our experiment, we conducted this process with 8K+ labeled data under the supervision of two human annotators.

\begin{table}[htbp]
\small
\centering
    \scalebox{.94}{
	\begin{tabular}{|l|l|c|c|c|} \hline
    \multirow{2}{*}{\textbf{Category}} & \multirow{2}{*}{\textbf{Features}} & \multicolumn{2}{c|}{\textbf{Class}} & \multirow{2}{*}{\textbf{Overall}}\\ \cline{3-4}
    &   &   \textbf{Normal} &   \textbf{WLT} &  \\ \hline
    Linguistic &    \# Posts &   8,421 &  255  & 8,676 \\ \hline
    \multirow{2}{*}{Visual}  &   \# Posts w/ images &  1,975 & 172  & 2,147\\ 
    &   \# Posts w/ OCR & 537 & 25  & 661\\ \hline
    \multirow{2}{*}{User} &  \# Users posted &  2,167 &  85 & 2,252\\ 
    & \# Users w/ profile &2,004&80&2,084 \\ \hline
    \multirow{3}{*}{Interaction} & \# Posts w/ URLs &  4,933  & 237  & 5,170\\
    &   \# Posts w/ hashtags & 2,740 & 149 & 2,889\\
    % &   \# Re-posts & 4,143 &  137 & 4,280\\
    &   \# Posts w/ mentions & 5,100 & 75 & 5,175\\ \hline
	\end{tabular}}
        \caption{WLT Dataset Info.}
        \label{tab:dataset}
\end{table}

\subsubsection{Dataset}
\label{sec:data_overall}
Table~\ref{tab:dataset} describes an overview of the collected dataset. We have 255 positive/WLT posts from 85 users and 8,421 hard negative/normal posts from 2,132 users. It is worth mentioning that the dataset is naturally multi-dimensional and multi-modal, where each post contains text, images, and social-platform-specific interaction tokens (e.g., links, hashtags, and mentions of other users). Another obvious observation is that although we only include hard negatives as normal posts, the dataset is still highly unbalanced regarding almost all reported categories and features.

\begin{figure}[tbp]
    \centering
    \includegraphics[width=.8\linewidth]{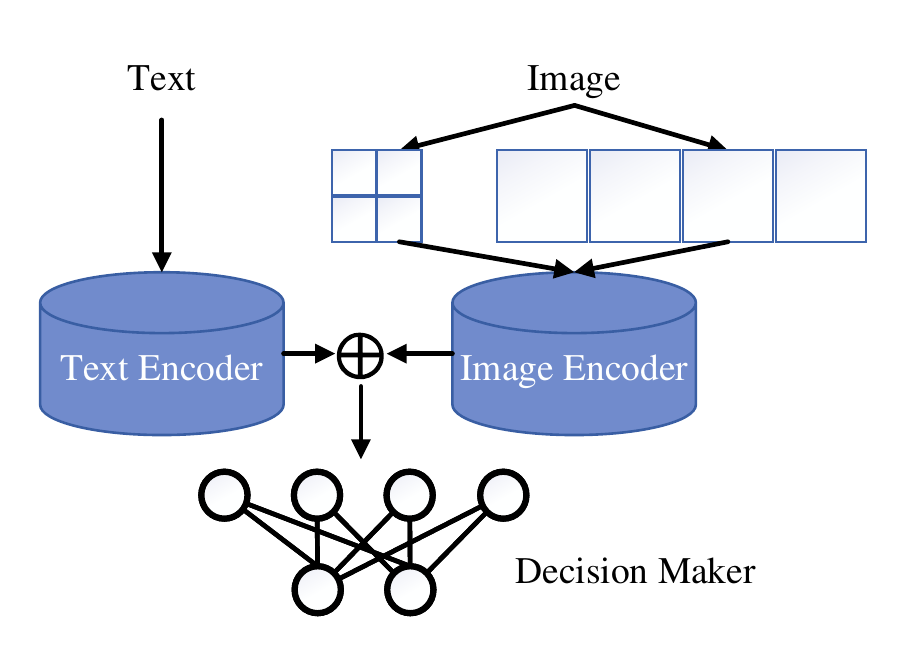}
    \caption{Visualizing the deep learning framework for human-in-the-loop labeling process.}
    \label{fig:framework}
\end{figure}

\begin{table*}[ht]
	\centering
    \small
    \scalebox{1.}{
	\begin{tabular}{|l|l|l|c|c|c|c|c|} \hline
    \multirow{2}{*}{\textbf{Modality}} & \multirow{2}{*}{\textbf{Model}} & \multirow{2}{*}{\textbf{Input}} & \multicolumn{2}{c|}{\textbf{WLT}} & \multicolumn{3}{c|}{\textbf{Overall}} \\ \cline{4-8}
    &&& \textbf{Pre.} & \textbf{Rec.} & \textbf{Macro F1} & \textbf{MCC} & \textbf{AUC} \\ \hline
    \multirow{7}{*}{Single-Modal} & Word Filter & \multirow{3}{*}{Text} 
        & $.460_{.000}$   & $\textbf{1.000}_{.000}$   & $.787_{.000}$   & $.641_{.000}$   & $.947_{.000}$\\ 
    & BERT & 
        & $.724_{.027}$ & $\underline{.942}_{.050}$ & $.899_{.005}$ & $.808_{.012}$ & $\textbf{.995}_{.003}$ \\ 
    & RoBERTa &
        & $.813_{.029}$ & $.835_{.031}$ & $.903_{.016}$ & $.807_{.032}$ & $.981_{.005}$\\ \cline{2-8}

    & ResNet& \multirow{2}{*}{$Image_S$} 
        & $.294_{.022}$ & $.846_{.094}$ & $.588_{.021}$  & $.335_{.063}$ & $.707_{.011}$  \\
    & ViT & 
        & $.268_{.051}$ & $.692_{.196}$ & $.569_{.047}$  & $.248_{.134}$ & $.696_{.020}$\\ \cline{2-8}
    & ResNet& \multirow{2}{*}{$Image_C$} 
        & $.322_{.046}$ & $.846_{.054}$ & $.616_{.049}$  & $.369_{.058}$ & $.798_{.032}$  \\
    & ViT & 
        & $.316_{.040}$ & $.862_{.148}$ & $.614_{.041}$  & $.373_{.101}$ & $.786_{.024}$ \\ \hline
    \multirow{2}{*}{Multi-Modal} & \multirow{1}{*}{BERT+ResNet} & \multirow{2}{*}{$Text + Image_C$} 
        & $\textbf{.878}_{.135}$ &  $.855_{.090}$ &  $\underline{.922}_{.023}$ & $\underline{.850}_{.045}$  & $\textbf{.995}_{.004}$ \\ 
    & \multirow{1}{*}{BERT+ViT}& 
        & $\underline{.821}_{.033}$  & $.928_{.090}$  & $\textbf{.929}_{.026}$  & $\textbf{.860}_{.053}$  & $\textbf{.994}_{.002}$\\ \hline
	\end{tabular}}
\caption{Experiment Results. We run each experiment three times and report the averages and the standard deviation (in underscript). In each column, best results are in bold and second best results are underlined.}
\label{tab:benchmark}
\end{table*}

\subsubsection{Model Design}
\label{sec:model_design}
Our automatic learning algorithm is a specialized deep learning framework designed for the task, as depicted in Fig.~\ref{fig:framework}. The framework comprises three major components: the \textit{text encoder}, the \textit{image encoder}, and the \textit{decision maker}. Given the multi-modality nature of the task, the model takes both text contents and corresponding images as inputs, producing a score between 0 and 1—an indicative probability of the input post being a WLT post.

\smallskip\noindent\underline{\textit{Text encoder:}} consumes the text content and generates its representation. We utilized pretrained language models such as BERT~\cite{devlin2019bert} and RoBERTa~\cite{liu2019roberta} as our backbone for text encoding.

\smallskip\noindent\underline{\textit{Image encoder:}} processes images and generates their corresponding representations. Our task, unique in handling varying numbers (0 to 4) of images per post, led us to two methods for addressing the multi-image problem: 1) Stitching: Down-sampling each image into 112 $\times$ 112 pixels and stitching the four images into one 224 $\times$ 224 image as input to the image encoder.
2) Concatenating: Using four images (without downsampling) as input to the image encoder, resulting in four distinct image representations. To achieve this, we employed pretrained vision models such as ResNet-50 \cite{he2016deep} and Vision Transformers (ViT) \cite{dosovitskiy2020image} as our backbone.

\smallskip\noindent\underline{\textit{Decision maker:}} fuses text and image representations to derive the final output. Specifically, the decision maker concatenates all representations and employs a multi-layer perceptron with layer dropouts and ReLU activations. While recognizing alternative options, such as attention mechanisms, for vector fusions, we leave these variations for future exploration, focusing on paving the way for further technical advancements in this work.

All components' parameters are finetuned and updated altogether with loss backpropagation. We adopt a typical cross-entropy classification loss for training.

\begin{table*}[htbp]
	\centering
    \small
    \scalebox{1.}{
	\begin{tabular}{|l|l|l|c|c|c|c|c|} \hline
    \multirow{2}{*}{\textbf{Modality}} & \multirow{2}{*}{\textbf{Model}} & \multirow{2}{*}{\textbf{Input}} & \multicolumn{2}{c|}{\textbf{WLT}} & \multicolumn{3}{c|}{\textbf{Overall}} \\ \cline{4-8}
    &&& \textbf{Pre.} & \textbf{Rec.} & \textbf{Macro F1} & \textbf{MCC} & \textbf{AUC} \\ \hline
    \multirow{6}{*}{Multi-Modal-$Image_{C}$} & \multirow{3}{*}{BERT+ResNet}& Text + Images 
        &  $.694_{.192}$ & $.814_{.123}$  &  $.846_{.045}$ & $.713_{.074}$  & $.975_{.007}$\\ 
    & &Text + Images + OCR 
        &  $\textbf{.878}_{.135}$ &  $.855_{.090}$ &  $\underline{.922}_{.023}$ & $\underline{.850}_{.045}$  & $\textbf{.995}_{.004}$\\ 
    & &Text + Images + OCR + Desc. 
        & $.816_{.058}$  &  $.754_{.050}$ & $.882_{.017}$  & $.765_{.033}$  & $.982_{.002}$\\ \cline{2-8}
    & \multirow{3}{*}{BERT+ViT}& Text + Images  
        &  $.810_{.165}$ & $.643_{.106}$  &  $.824_{.024}$ & $.678_{.034}$  & $.977_{.008}$\\ 
    & &Text + Images + OCR  
        & $.821_{.033}$  & $\underline{.928}_{.090}$  & $\textbf{.929}_{.026}$  & $\textbf{.860}_{.053}$  & $\textbf{.994}_{.002}$\\ 
    & &Text + Images + OCR + Desc. 
        & $.686_{.206}$  &  $.725_{.176}$ &  $.819_{.004}$ & $.659_{.015}$  & $.973_{.014}$\\ \hline

    \multirow{6}{*}{Multi-Modal-$Image_{S}$} & \multirow{3}{*}{BERT+ResNet} & Text + Images 
        &  $.782_{.050}$ & $.783_{.075}$  &  $.881_{.034}$ & $.763_{.067}$  &$.982_{.005}$\\ 
    & &Text + Images + OCR  
        &  $.763_{.148}$ & $.812_{.181}$  & $.871_{.011}$  &  $.756_{.021}$ &$.991_{.002}$\\ 
    & &Text + Images + OCR + Desc. 
        & $.668_{.068}$  & $\textbf{.957}_{.075}$  & $.879_{.015}$  & $.777_{.019}$  &$.990_{.009}$\\ \cline{2-8}
     & \multirow{3}{*}{BERT+ViT} & Text + Images  
         &  $.803_{.130}$ & $.806_{.035}$  & $.886_{.051}$  & $.780_{.093}$  &$.983_{.013}$\\ 
    & &Text + Images + OCR   
        &  $\underline{.854}_{.052}$ & $.872_{.110}$  & $.921_{.040}$  & $.847_{.073}$  &$.993_{.005}$\\ 
    & &Text + Images + OCR + Desc.  
        &  $.732_{.257}$ & $.725_{.153}$  & $.830_{.036}$  & $.682_{.064}$  &$.980_{.011}$\\ \hline
	\end{tabular}}
    \caption{Detailed Experiment Results.}
    \label{tab:variation}
\end{table*}

\section{Benchmark Results}
\label{sec:benchmark}
% We illustrate our experiment details and results in the following subsections.

\subsection{Models}
We consider the following models as experiment baselines:

\subsubsection{Single Modality Models:}
We first experiment with single-modality models, including text-only models that consume only text as inputs, and image-only models that consume only images as inputs.

\smallskip\noindent\textit{Text-Only Models:}
We mainly used existing pretrained language models and finetune them on our dataset for downstream classification task. We also introduce a naive method serving as a lower-bound baseline due to its intuitively straightforward implementation.
\squishlist
\item \textbf{Word Filter:} This naive method consumes the lowercased post text and only looks for the word ``ivory''. If the text contains such a word, the method will predict it as a WLT post. The naive approach is a lower bound for its straightforward design and low computation cost.
\item \textbf{BERT:} We finetune a BERT~\cite{devlin2019bert} on the dataset. While training large language models (LLMs) is too costly for many researchers, finetuning BERT is common and practical. More specifically, we finetuned the pretrained ``BERT-base-uncased''.
\item \textbf{RoBERTa:} We also finetune RoBERTa~\cite{liu2019roberta} which claims to have a more robust performance in general. Specifically, we finetuned the pretrained ``RoBERTa-base-uncased''.
\squishend

\smallskip\noindent\textit{Image-Only Models:}
Similarly, we finetune two existing pretrained models in the computer vision (CV) domain, including convolutional-based methods and transformer-based methods. It is worth noting that the number of images per post can vary from zero to four, adding complexity to the research. For each tweet, we process the data in a padding-with-empty way, meaning that regardless of whether the tweet has attached images, we always create four placeholders for the images. We use all $0$ to fill in the placeholders if there are insufficient images, resulting in a $3 \times 224 \times 224$ image for each tweet.

\squishlist
\item \textbf{ResNet-50:} Convolutional neural networks (CNNs) have achieved remarkable success in image recognition tasks. We adopt the pretrained ResNet-50~\cite{he2016deep}, widely used due to its transferability. 
\item \textbf{ViT:} With the development of transformers in NLP domain~\cite{vaswani2017attention}, Vision Transformers (ViT) have emerged as a groundbreaking architecture~\cite{dosovitskiy2020image}. ViTs break down an image into fixed-size patches and linearly embed each patch. We applied the pretrained ``ViT-base'' for finetuning.
\squishend

\smallskip\noindent\textit{Bi-Modality Models:}
We experiment with variations of our design illustrated in Sec.~\ref{sec:model_design} for bi-modality as below:
\squishlist
\item \textbf{Image processing variants:} $image_S$ stands for image stitching method while $image_C$ accounts for image flattening/concatenation method.
\item \textbf{Text processing variants:} We had pure text input, text+OCR (i.e., post text and text extracted from images) results as text input, and text+OCR+user profile description as text input variants.
\squishend

\subsection{Experiment Settings}
To address the dataset imbalance, we down-sample the negative class to keep the class ratio between WLT and normal posts as 1:10 (i.e., We sample 2,550 normal posts from the complete set of negative class). We extract all posts containing ``ivory'', then randomly sample the rest of the instances to meet the requirement. The data is split into train/dev/test sets (70/20/10 ratio), with the constraint that posts from the same user can only show up in at most one split. We then combine the corresponding splits of both classes to form the final data split. The intuition behind such a splitting scheme is: 1) within each class, we prevent user information leakage. 2) On the contrary, we allow posts from the same user in different classes to show up in different splits, as a user can post WLT as well as normal ones. We train/finetune models on the training set using various hyperparameter combinations, selecting the best-performing model on the dev set for evaluation on the test set. All models are run on machines with Nvidia GeForce 3090 GPUs.

\subsection{Evaluation Metrics}
We evaluate model performance with these metrics: 
\squishlist
\item For the concerned positive class (WLT), we measure the precision (Pre.) and recall (Rec.) rate.
\item For the overall evaluation, we measure the Macro F1, Matthews Correlation Coefficiency (MCC), and Area Under the Curve score (AUC).
\squishend
We run each model with best-performing hyperparameters under three random seeds and report their metric results with average scores and standard deviation. As the data is still imbalanced, we mitigated the issue with inverse class weights, and then we recalibrated the classification threshold towards maximizing the MCC score on the validation set.

\subsection{Main Results}
\label{sec:exp_result}
Table~\ref{tab:benchmark} presents the results of our experiments, and noteworthy observations are highlighted below:

\noindent\underline{\textit{Naive method has shortcomings despite some merit:}} 
The naive method (word filtering) achieved a perfect positive class recall rate and a relatively high AUC. However, its limitation lies in its inability to recognize positive posts lacking the keyword, leading to misclassifications of negative posts containing the keyword. This results in lower Macro F1 and MCC scores, emphasizing the constraints of word filtering without considering post context. It's essential to note that the ``Word Filter'' method is deterministic, resulting in results with a standard deviation of 0.

\noindent\underline{\textit{Text modality has an edge over vision:}} 
In the comparison between text-based and image-based single-modal models, text-based models exhibited superior positive class precision and overall performance. This advantage is partially attributed to many samples lacking images in the original posts (see Fig.\ref{fig:image_num} and Sec.\ref{sec:image_stat} for details). Despite this, visual models demonstrated merit, especially in multi-modal settings, as discussed in the following paragraph. Additionally, ViT consistently outperformed ResNet under both image concatenation and image stitching settings. Furthermore, each model generally performed better with image concatenation than image stitching.

\noindent\underline{\textit{Multi-modality models achieve overall best results:}} 
Moreover, we list the best-performing results for our bi-modal variations. For the bi-modality models, we observe the best precision result for the concerned positive class (0.878 for BERT+ResNet and 0.821 for BERT+ViT). Meanwhile, we also observe the best performance for BERT+ViT across all three overall metrics and BERT+ResNet achieves similar/second best performance. The best multi-modal result (BERT+ViT) gained an improvement against the best single-modal results of 2.9\% on Macro F1 and 6.4\% on MCC. Such a strong result indicates embracing the advantage from both modalities effectively improves the performance under various metrics.

\subsection{Multi-modal Variation Results}
We conducted additional experiments exploring various modeling possibilities for multi-modal frameworks, including image OCR (Optical Character Recognition), user descriptions, and their combinations with the previously mentioned model variations in Table~\ref{tab:variation}. Key findings include:

\squishlist
\item OCR proved beneficial in extracting relevant information embedded in images, such as product descriptions, sales prices, and selling websites. This enhancement contributed to overall improved performance.
\item The inclusion of user descriptions had a negative impact on model performance, potentially due to the limited number of users. With fewer posting users than posts, the model tended to overfit on user attributes.
\item Similar to the single modality results, multi-modal results w/ ViT consistently outperform those w/ ResNet.
\squishend
These benchmarking results and insights lay the foundation for further improvements in classification results, which can be explored in future work.

\begin{figure*}[ht]
    \centering
    \small
    \subfloat[Post length w.r.t \# tokens]{
        \includegraphics[width=.5\columnwidth]{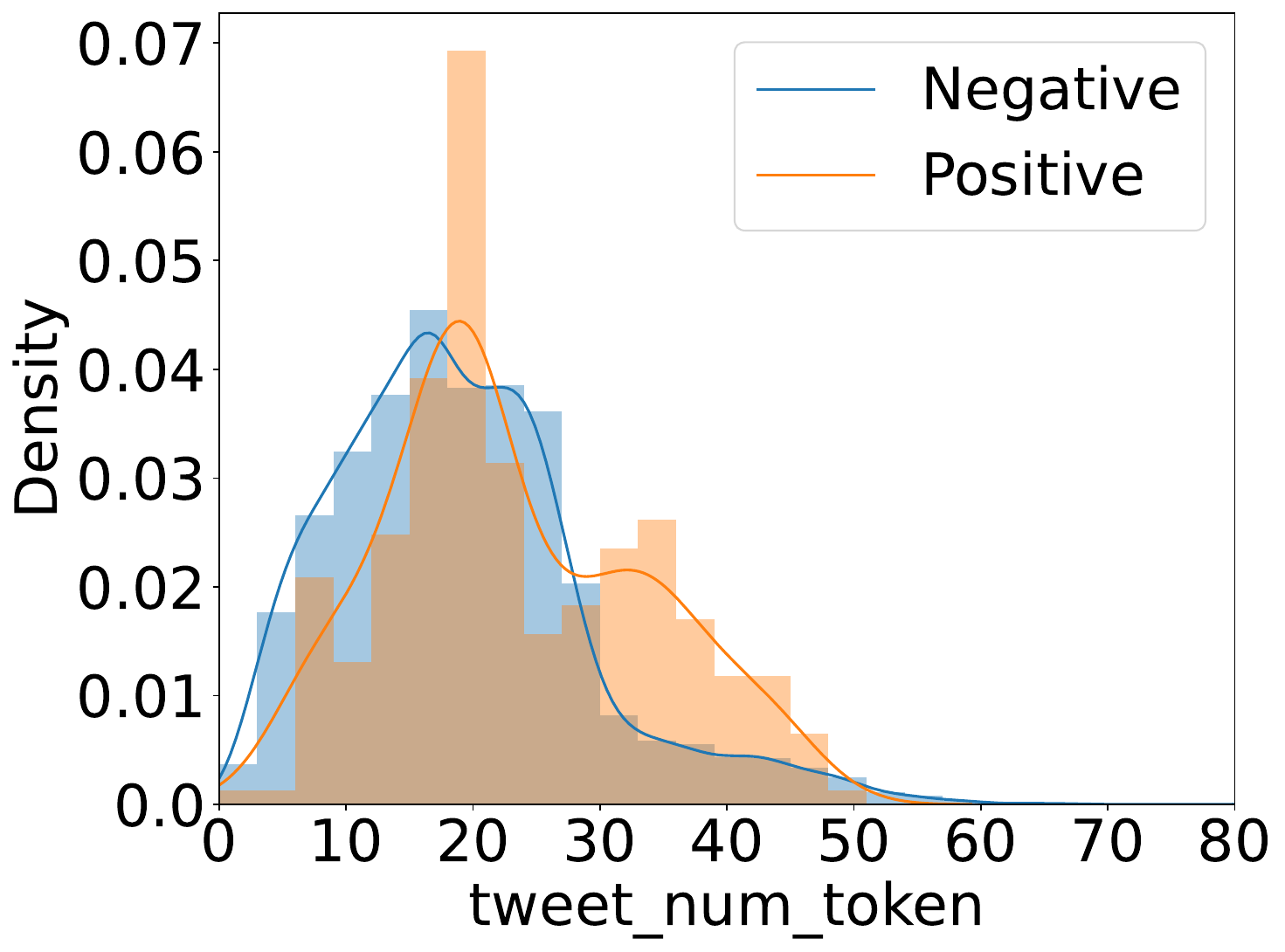}
        \label{fig:dis_len_token}
    }
    % \hspace{5pt}
    \subfloat[Post length w.r.t \# chars]{
        \includegraphics[width=.5\columnwidth]{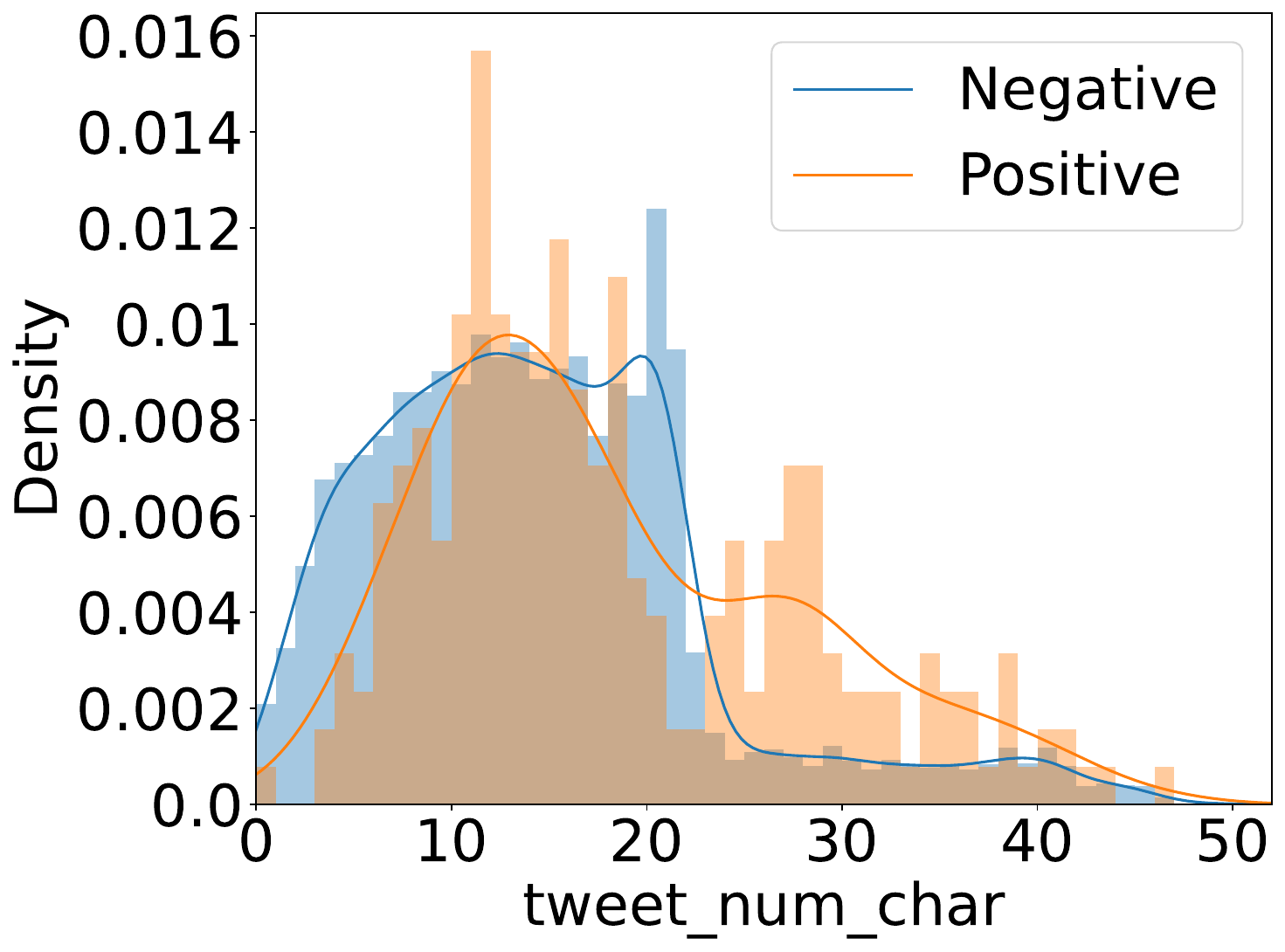}
        \label{fig:dist_len_char}
    }
    \subfloat[Avg. token length in \#chars.]{
        \includegraphics[width=.5\columnwidth]{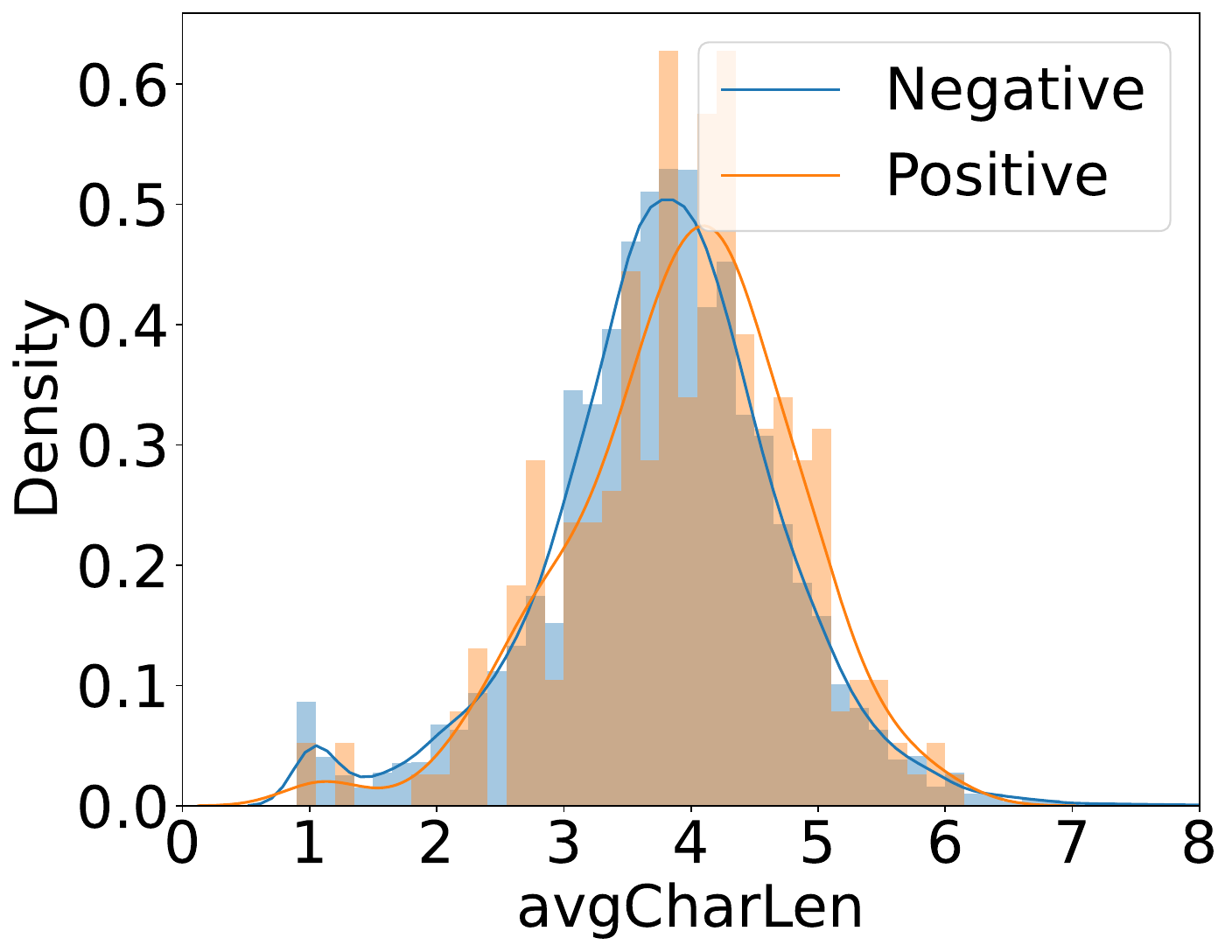}
        \label{fig:dis_token_in_char}
    }
    % \hspace{5pt}
    \subfloat[Avg. token length in \#chars w/o stop words.]{
        \includegraphics[width=.5\columnwidth]{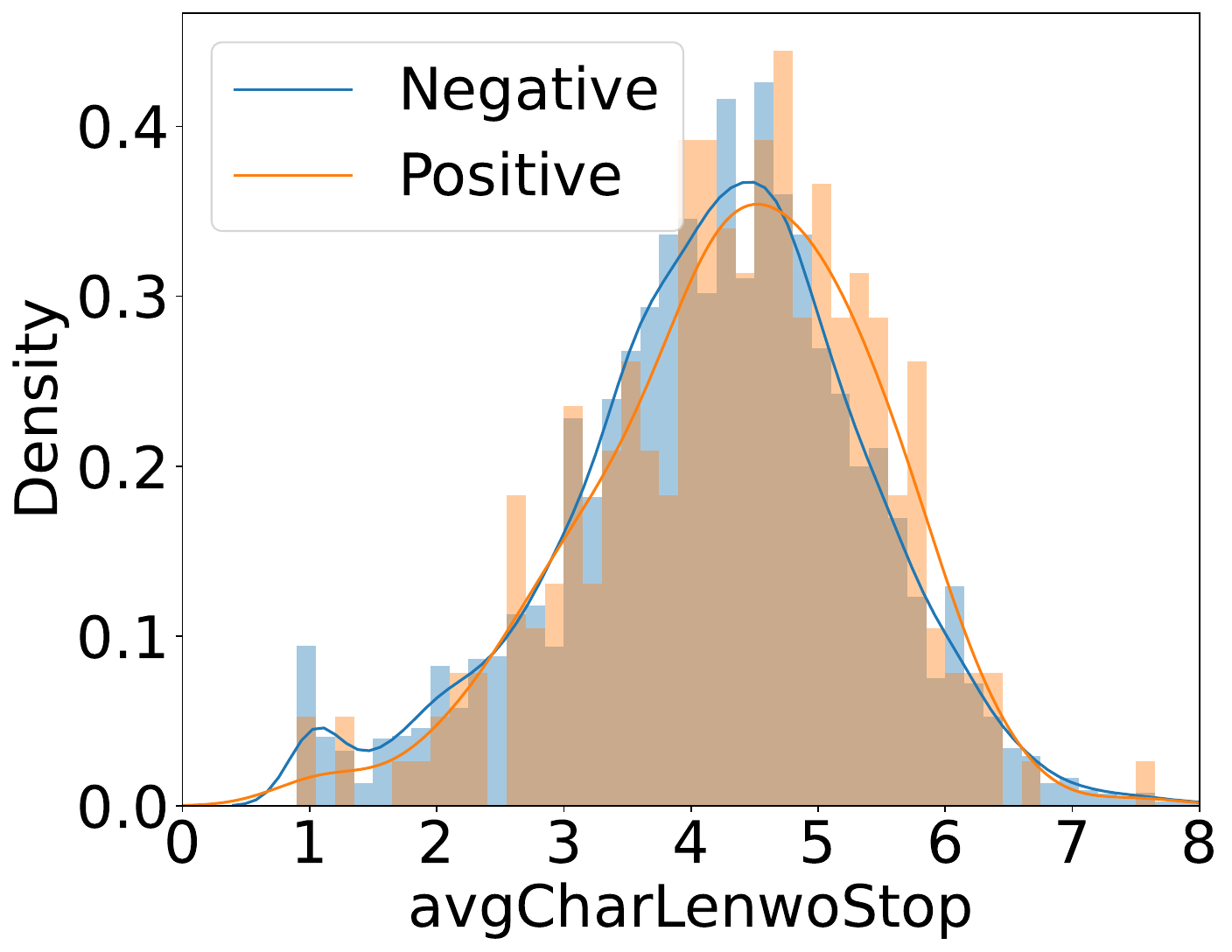}
        \label{fig:dist_token_in_char_nonstop}
    }
    % \hspace{5pt}
    \caption{Distribution of text length for (a) positive and (b) negative posts, as well as average token length (c) w/ and (d) w/o stop words. It is worth noting that normal posts tend to have long tails on the post length and average token length, but we had cutoffs in the visualizations to focus on the left parts. For more information on the max length, please refer to Table~\ref{tab:text_stat}.}
    \label{fig:tweet_length}

\end{figure*}

\section{Dataset Analysis}
\label{sec:dataset}
This section unveils intriguing insights gleaned from the labeled dataset, as showcased in Table~\ref{tab:dataset}. We present our findings from the analysis of textual and visual modalities in Sec.~\ref{sec:data_text} and~\ref{sec:data_image}, respectively. Additionally, we delve into the distributions of OSN-specific tokens, such as mentions, URLs, and hashtags, in Sec.~\ref{sec:data_other}. While our preliminary experiments primarily focused on harnessing deep learning methods for classification, these observations serve as valuable guidance for future research endeavors, particularly in the development of essential handcrafted features for general machine learning frameworks.

\begin{table}[htbp]
	\centering
    \small
    \scalebox{.98}{
	\begin{tabular}{|l|c|c|c|c|c|c|} \hline
    \multirow{2}{*}{\textbf{Category}} &     \multicolumn{3}{c|}{WLT} &  \multicolumn{3}{c|}{Normal} \\ \cline{2-7}
    &   avg. &  std. &  max  &   avg. &  std. &  max \\ \hline
    \#words &   23  & 10.3  & 49    & 19    & 10    & 78 \\ 
    \#words w/o ST  & 18    & 10    & 47    & 16    & 10    & 78 \\ \hline
    \#chars &  91   & 48    & 232   & 72    & 43    & 246 \\ 
    \#chars w/o ST  & 86    & 48    & 231   & 69    & 43    & 244 \\ \hline
    \#char/word   & 3.9   & 0.9   & 6.1   & 3.8   & 1.0   & 18 \\ 
    \#char/non-ST & 5.1   & 1.1   & 11  & 4.9   & 1.6   & 18 \\ 
    \#char/non-SW & 4.3   & 1.1   & 7.5   & 4.2   & 1.2   & 18 \\ 
    \#char/non-SW/ST  & 5.9   & 1.3   & 11  & 5.7   & 1.9   & 18 \\ \hline
	\end{tabular}}
     \caption{Dataset Text Statistics. ST: special tokens (retweet/rt, mentions, hashtags, urls, etc.). SW: stop word.}
     \label{tab:text_stat}
\end{table}

\begin{figure}[ht]
    \centering
    \subfloat[Sentiment Analysis for posts. ]{
        \includegraphics[width=.8\linewidth]{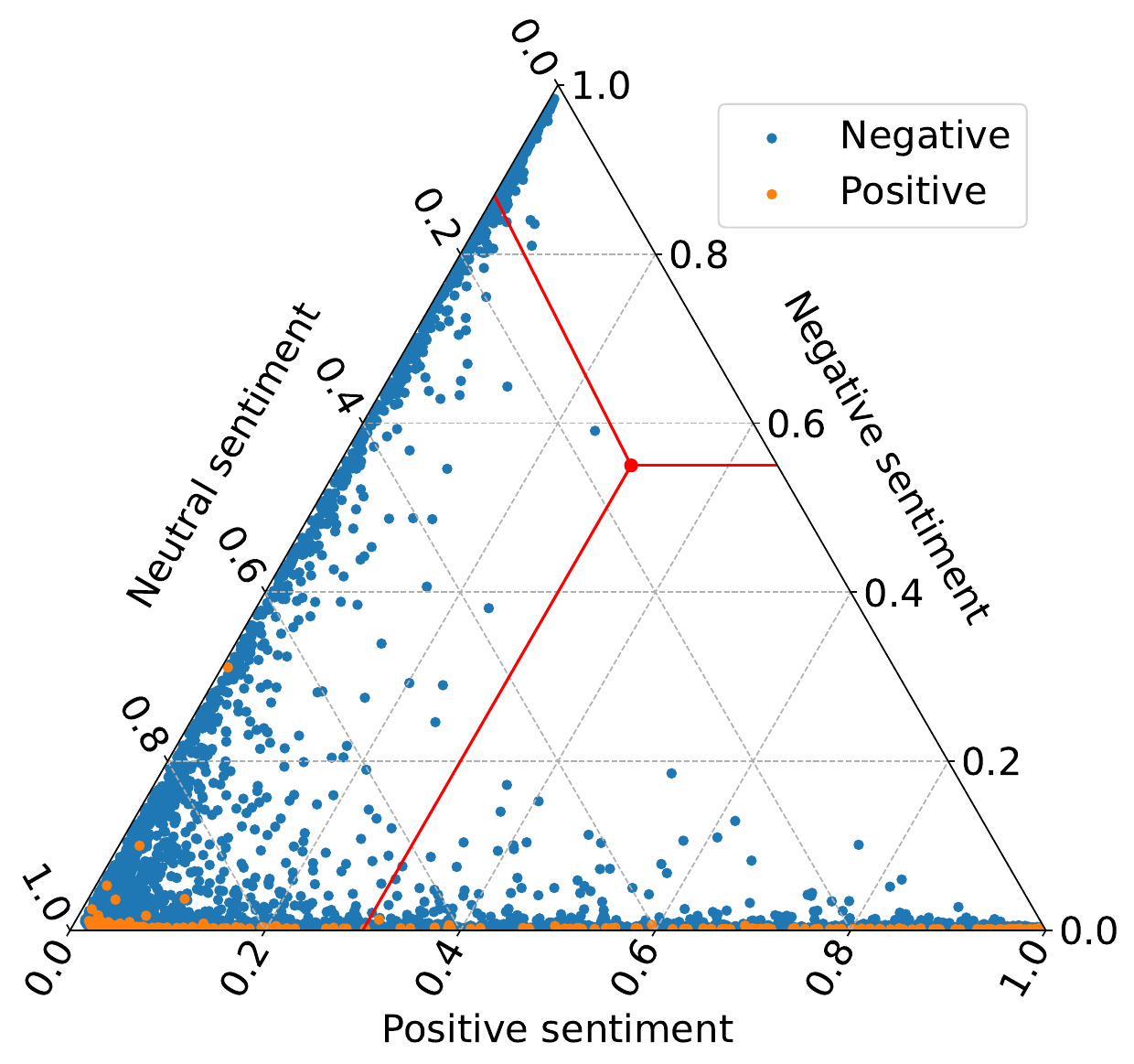}
        \label{fig:sentiment}
    } \\
    \subfloat[Toxicity Analysis for posts.]{
        \includegraphics[width=.97\linewidth]{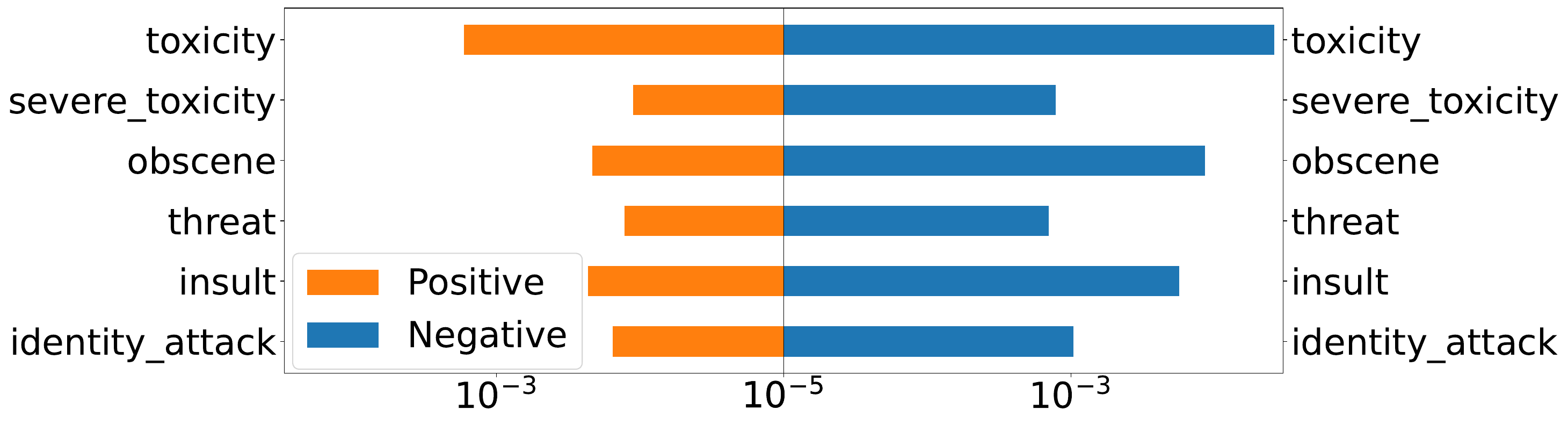}
        \label{fig:toxicity}
    }
    \caption{Semantic Analysis for posts.}
    \label{fig:semantics}
\end{figure}

\begin{figure}[ht]
    \centering
    \subfloat{
        \includegraphics[width=.75\linewidth]{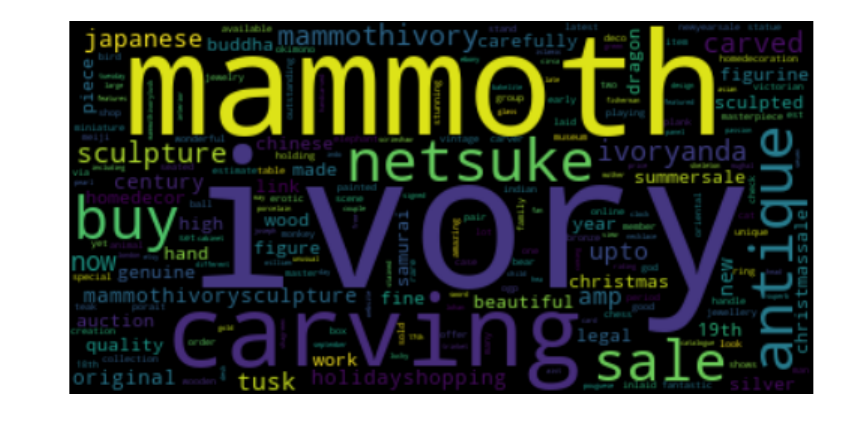}
        \label{fig:pos_wc}
    }
    \
    \subfloat{
        \includegraphics[width=.75\linewidth]{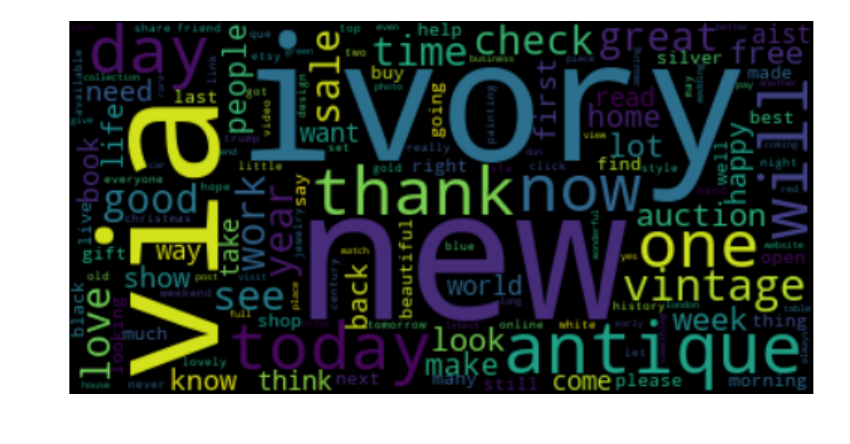}
        \label{fig:neg_wc}
    }
    \caption{Word cloud for pos.(top) \& neg.(bottom) posts.}
    \label{fig:wordcloud}
\end{figure}

\subsection{Text Analysis}
\label{sec:data_text}
We present comprehensive text analysis statistics in Table~\ref{tab:text_stat}. Additionally, detailed distributions of text features within three major categories: writing styles, text quality, and text semantics, are illustrated in finer granularity through figures in the following subsections.

\subsubsection{Writing Style}
\label{sec:text_style}
WLT posts differ from normal posts in their intention, reflecting the variance in writing styles. We analyze the writing style differences concerning post length and average word length. NLTK~\cite{loper2002nltk} Tweet tokenizer is employed for tokenizing text into words.

\textit{Post Length:}
The distribution of text posts is presented in Table~\ref{tab:text_stat} and Fig.~\ref{fig:tweet_length} from two perspectives: post length in terms of the number of tokens and characters. Generally, positive posts are longer on average, typically displaying two local maxima in distributions. Negative samples, while having a lower average post length, exhibit a long tail accounting for more extreme cases. This difference results in a wider range of text lengths for negative samples.

\textit{Word Length:}
Average token/word length in the number of characters per post is considered. The results are shown in Table~\ref{tab:text_stat}. Positive posts tend to use longer tokens, but negative samples have a long tail for extreme cases. The impact of stop words and unique tokens on OSNs is showcased, with minimal influence on the relative comparison results. Distribution visualizations in Fig.~\ref{fig:tweet_length} support these findings.

\subsubsection{Text Semantics}
\label{sec:sentiment_toxicity}
We investigate the nature of text semantic differences between the two classes, focusing on sentiments and toxicity levels in Fig. \ref{fig:sentiment} and Fig. \ref{fig:toxicity}.

\textit{Sentiment Analysis:}
VADER~\cite{hutto2014vader} is used for evaluating text sentiments. Each dot in Fig.~\ref{fig:sentiment} represents a post with three-dimensional scores (positive, neutral, and negative sentiment score). WLT posts predominantly exhibit positive/neutral sentiments, while normal posts display a more even distribution of all sentiments.

\textit{Toxicity Analysis:}
The pretrained deep learning model DeToxifying~\cite{bose2023detoxifying} measures text toxicity levels. Average toxicity scores of WLT posts and normal posts are compared in Fig.~\ref{fig:toxicity}. Both posts mostly have relatively low toxicity across all categories, with WLT posts showing even lower toxicity scores due to their product-promoting intentions.

\subsubsection{Frequent Word Usage}
We conduct an analysis of the most commonly used words in both classes, illustrating word clouds of WLT and normal posts in Fig.~\ref{fig:wordcloud}, where larger fonts signify higher occurrence within the dataset. The presence of shared frequent words such as "ivory" underscores the constraints of simplistic keyword filtering methods. Further validation in the experiment and case study sections yields both quantitative and qualitative results, enriching our understanding of the efficacy of our approach.

\begin{figure}[hbp]
    \centering
    \small
    \subfloat[Flesch Reading Ease]{
        \includegraphics[width=.48\columnwidth]{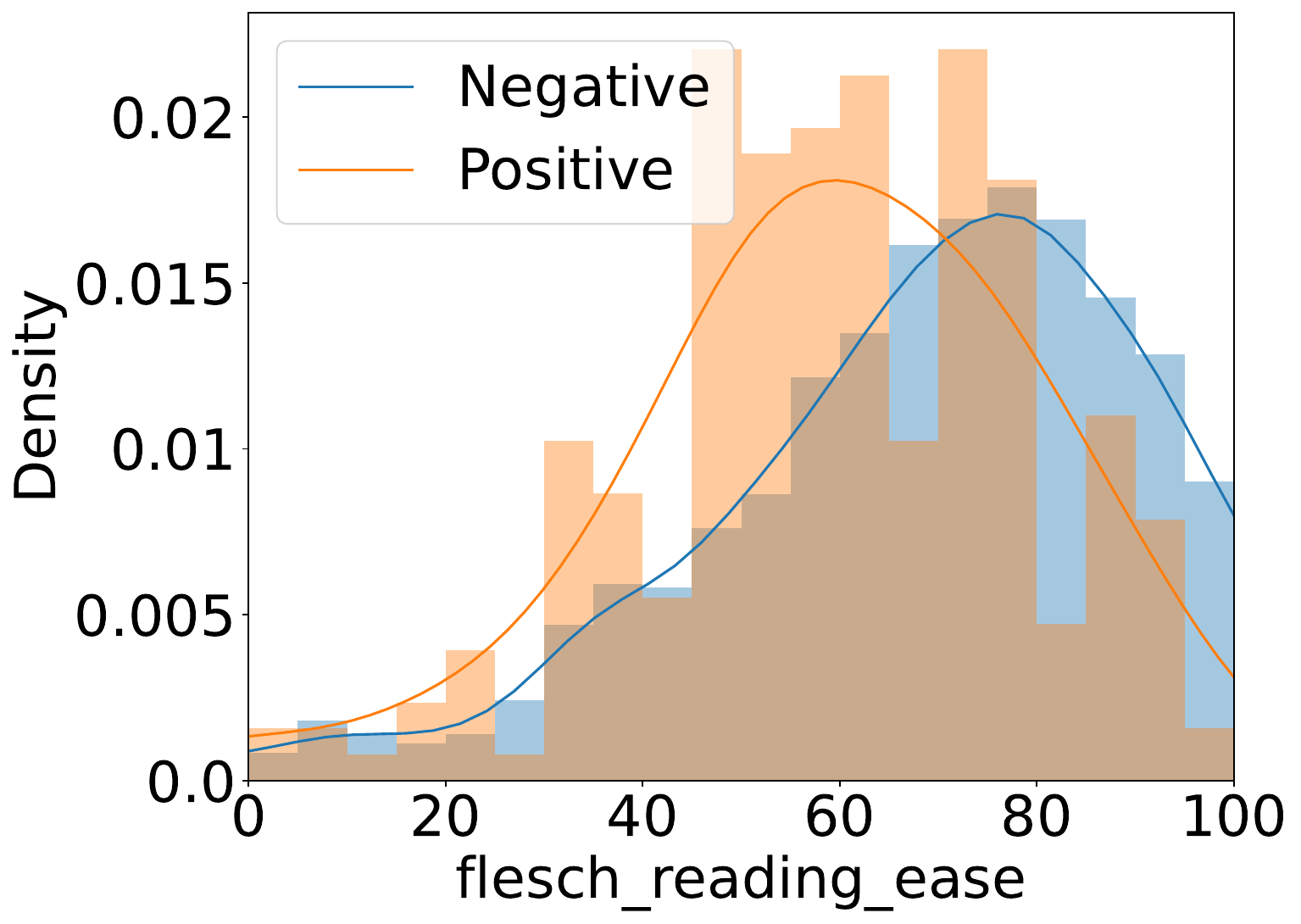}
        \label{fig:readability}
    }
    \subfloat[CoLA score.]{
        \includegraphics[width=.46\columnwidth]{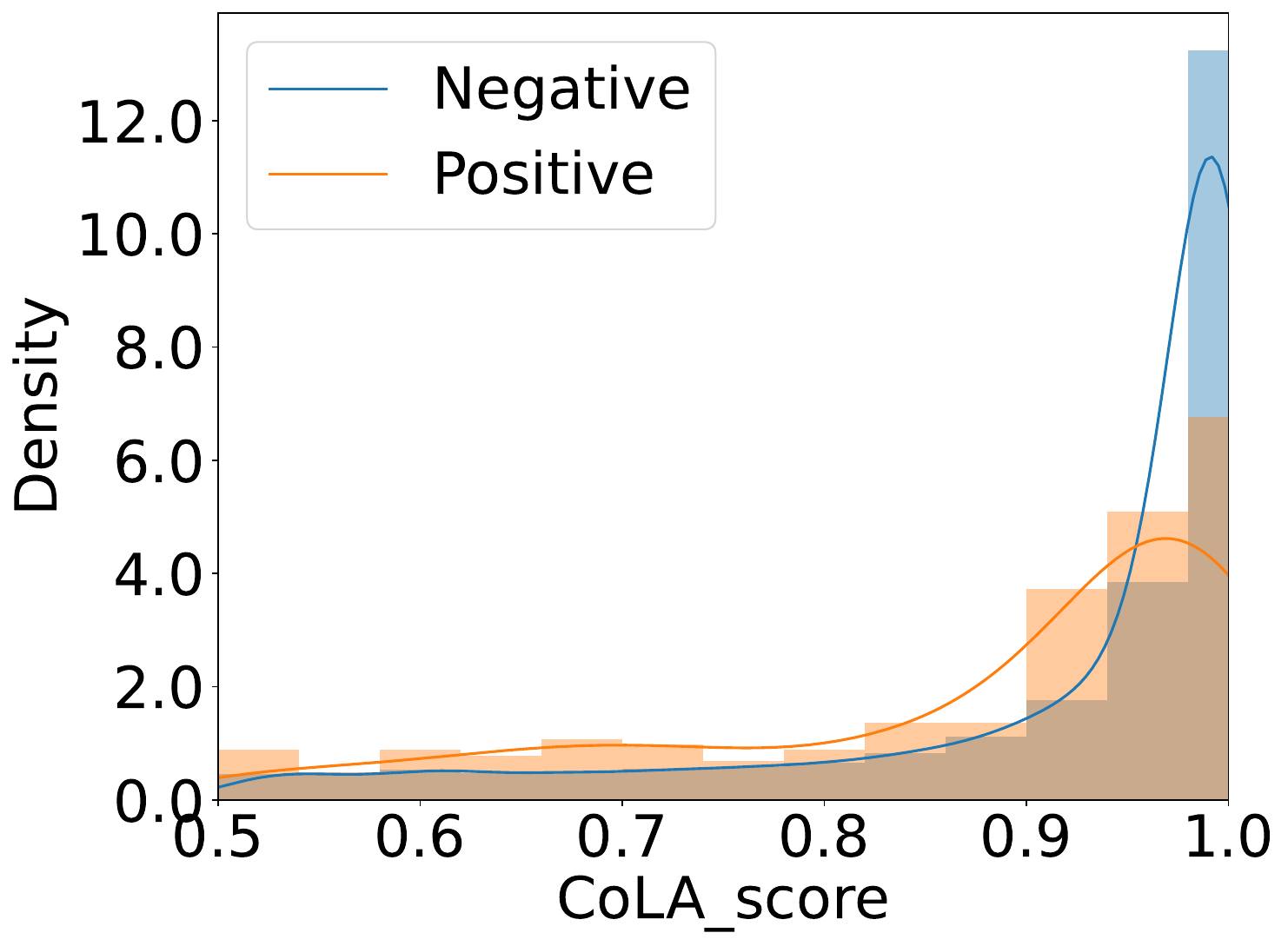}
        \label{fig:linguistic_acceptability}
    }
    \caption{Text Readability and Linguistic Acceptability.}
    \label{fig:text_quality}
\end{figure}

\begin{figure}[hbp]
    \centering
    \includegraphics[width=.62\linewidth]{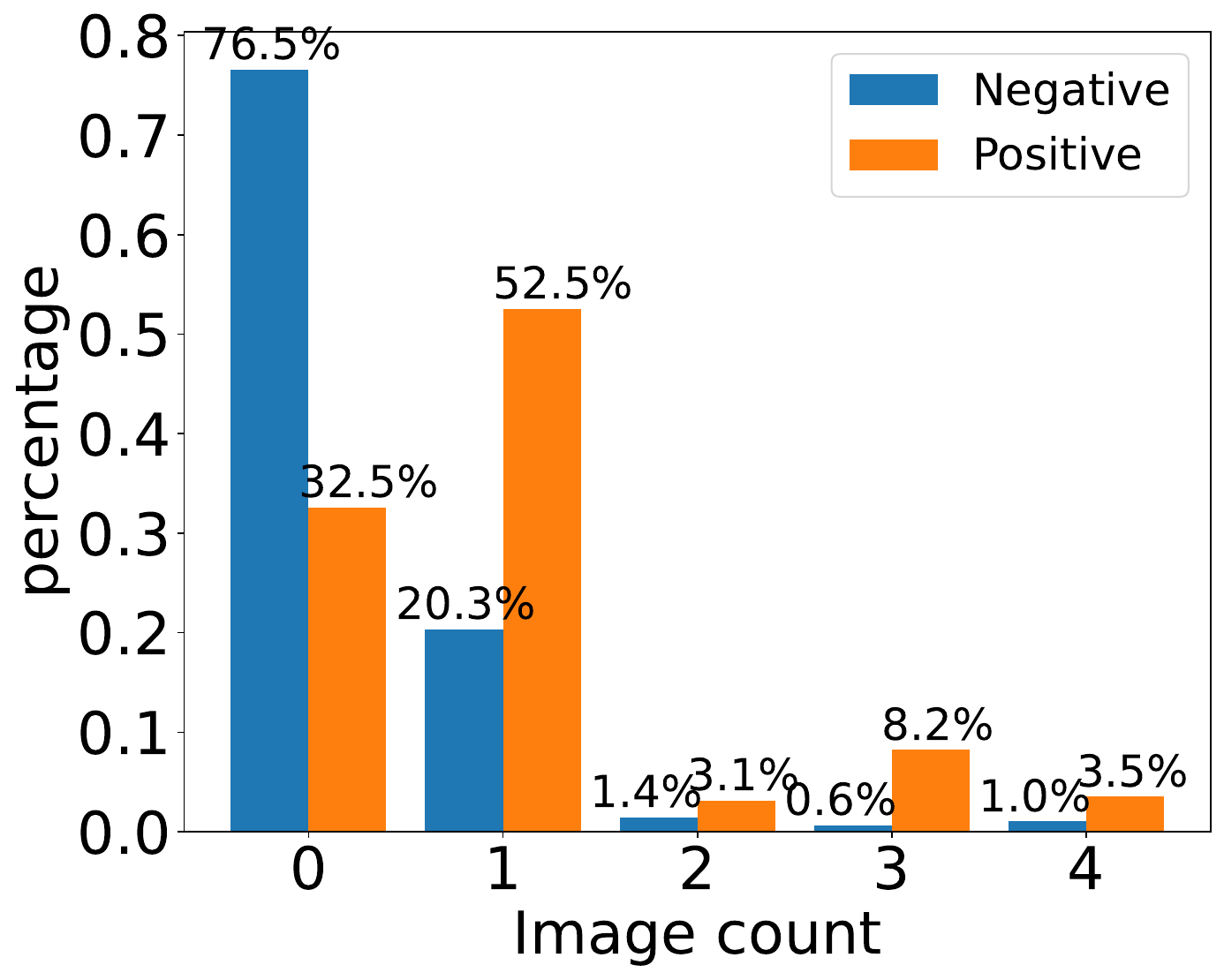}
    \caption{Distribution of \# images in posts.}
    \label{fig:image_num}
\end{figure}

\begin{figure*}[htbp]
    \centering
    \small
    \subfloat[\# Hashtag Distribution]{
        \includegraphics[width=.27\linewidth]{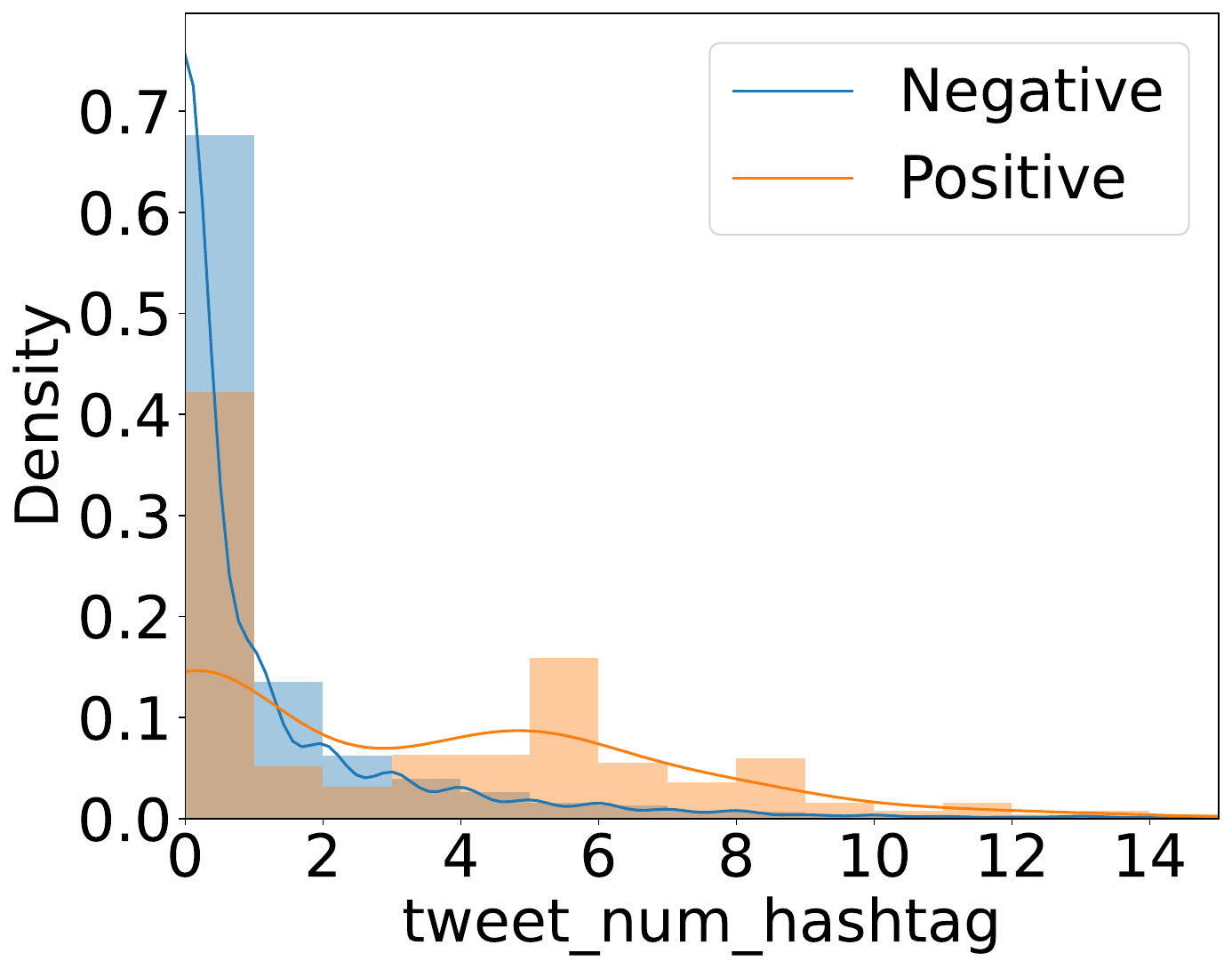}
        \label{fig:hashtag_dist}
    }
    \hspace{5pt}
    \subfloat[\# Mention Distribution]{
        \includegraphics[width=.27\linewidth]{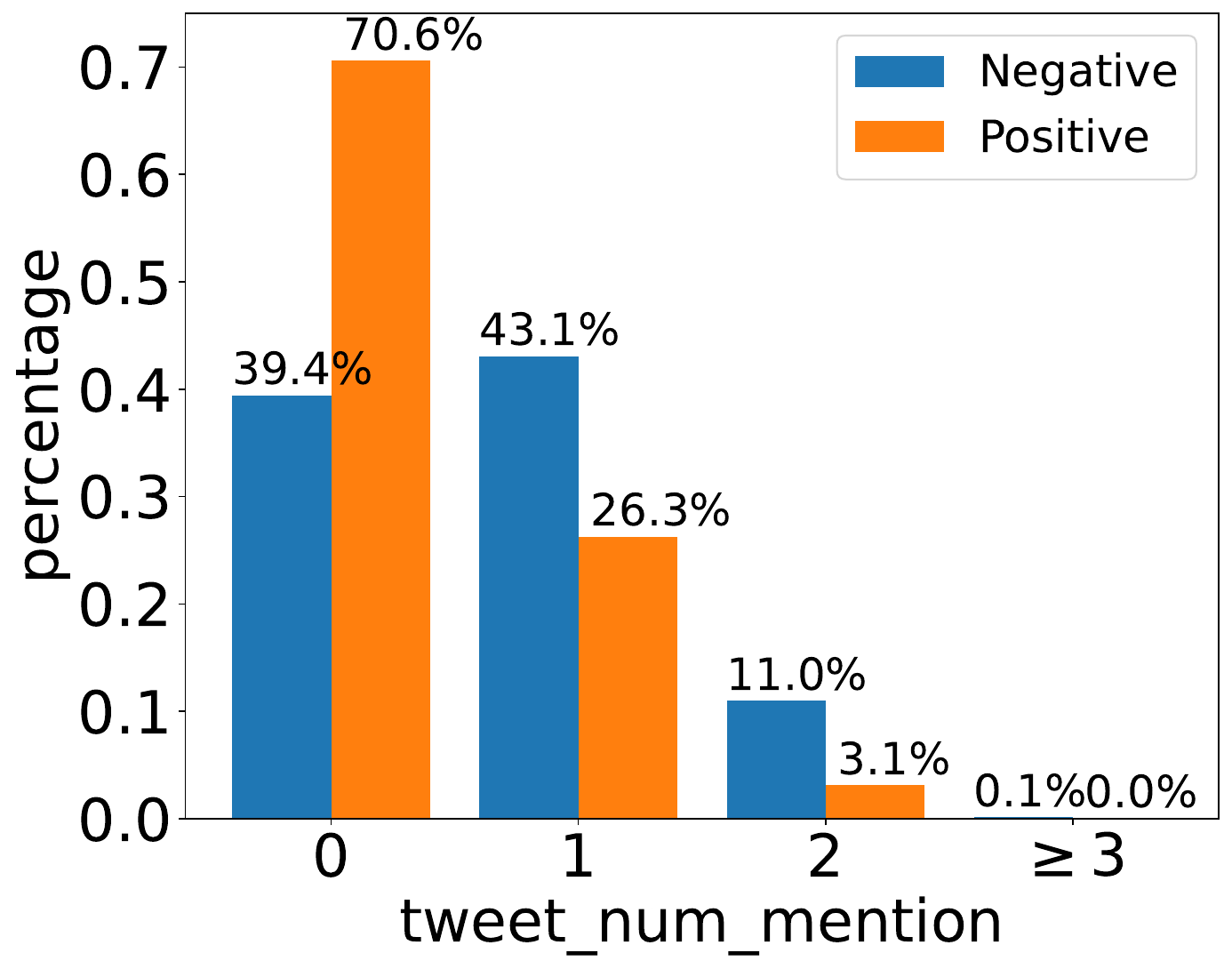}
        \label{fig:mention_dist}
    }
    \hspace{5pt}
    \subfloat[\# URL Distribution]{
        \includegraphics[width=.285\linewidth]{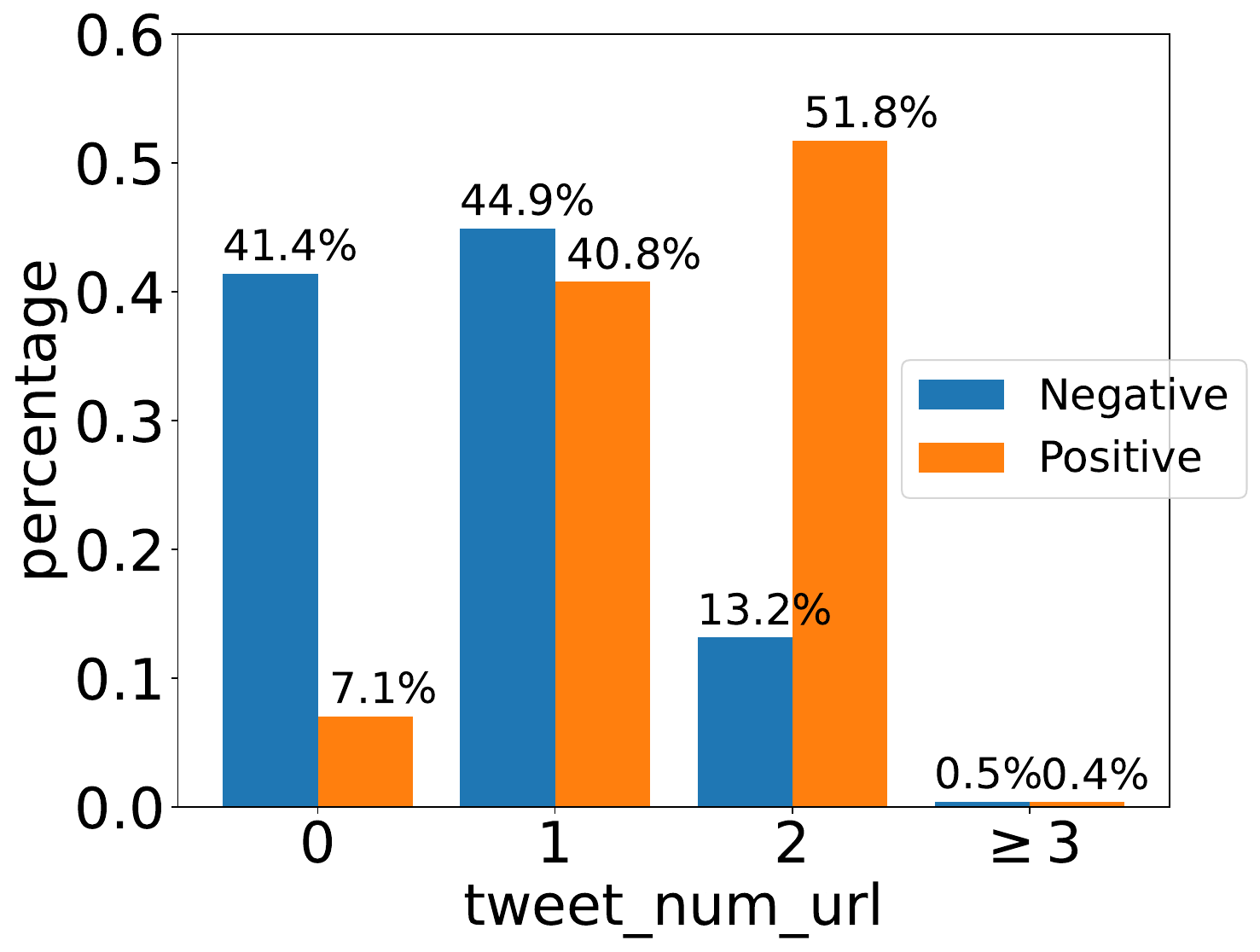}
        \label{fig:url_dist}
    }
    \caption{Distribution of special tokens. Note that we truncate Fig.~\ref{fig:hashtag_dist} positive class tails for better visualization.}
    \label{fig:special_token}
\end{figure*}

\subsubsection{Text Quality}
\label{sec:text_quality}
Another important perspective for the posts is their text quality. We analyze the posts' language quality in two ways: the text readability and the linguistic acceptability. More specifically, 
\squishlist
\item We measure text readability with Flesch Reading Ease Score~\cite{flesch1979write}, computed according to Eq.~\ref{eq:reading_ease}.
\begin{equation}
% \small
    s = 206.835 - 1.015(\frac{words}{sentences}) - 84.6(\frac{syllables}{words})
    \label{eq:reading_ease}
\end{equation}
The Flesch Reading Ease typically ranges between 0 and 100, where a higher score means the text is easier to read.
\item We leverage a language model BERT~\cite{devlin2019bert} finetuned on the CoLA~\cite{warstadt2018neural} dataset, specifically focused on predicting language acceptability.\footnote{\small\url{https://tinyurl.com/tcdnj59a}} The finetuned model will infer on our data, giving each sample a score between 0 and 1, where a higher score indicates higher linguistic acceptability. 0 means not acceptable, while 1 means perfectly acceptable text in the linguistic sense.
\squishend
The distribution results are shown in Fig.~\ref{fig:text_quality}. We noticed that: 1) the majority of the posts, regardless of their labels, are within decent scores for both readability and linguistic acceptability; 2) The positive posts tend to have lower readability scores and linguistic acceptability scores;
These observations may be because positive posts are more prolonged and use more complicated words (thus less readability and linguistic acceptability intuitively).

\begin{table}[htbp]
\small
	\centering
    \scalebox{1.}{
	\begin{tabular}{|l|c|c|c|c|c|c|} \hline
    \multirow{2}{*}{\textbf{Category}} &     \multicolumn{3}{c|}{WLT} &  \multicolumn{3}{c|}{Normal} \\ \cline{2-7}
    &   avg. &  std. &  max &   avg. &  std. &  max \\ \hline
    \# URLs &1.5&0.6&3&0.7&0.7&4 \\ 
    \# mentions &0.3&0.5&2&0.96&1.7&50 \\
    \# hashtags &3.3&3.9&23&1.0&2.2&24 \\
    % \# reposts &&&&&&&& \\ 
    \hline
	\end{tabular}}
 \caption{Dataset Special Token (ST) Statistics.}
 \label{tab:special_token_stat}
\end{table}
\subsection{Interactions and Special Activities}
\label{sec:data_other}
Special tokens' statistics are provided in Table~\ref{tab:special_token_stat}, with detailed distributions presented in Fig.~\ref{fig:special_token}. Special tokens include links, mentions, and domain-specific hashtags. WLT posts tend to have more links and hashtags for product promotion purposes but fewer mentions and reposts, aligning with the expectation of broadcasting to a wider audience rather than specific user notifications.

\subsection{Image Analysis}
\label{sec:data_image}
In the following subsections, we perform image analysis on images in WLT (positive) posts and normal (negative) posts.  

\subsubsection{Image Number Statistics}
\label{sec:image_stat}
Our selected $8,676$ tweets contain 2,713 images and 2,147 tweets have at least one image as shown in Table~\ref{tab:dataset} . Detailed distribution is shown in Fig.~\ref{fig:image_num}, showing that 76.5\% normal posts (negative class) do not contain images. On the other hand, WLT posts (positive class) consistently have more images (1, 2, 3, and 4 images).

\begin{figure}[ht]
    \raggedleft
    \small
    \subfloat{
        \includegraphics[width=.90\linewidth]{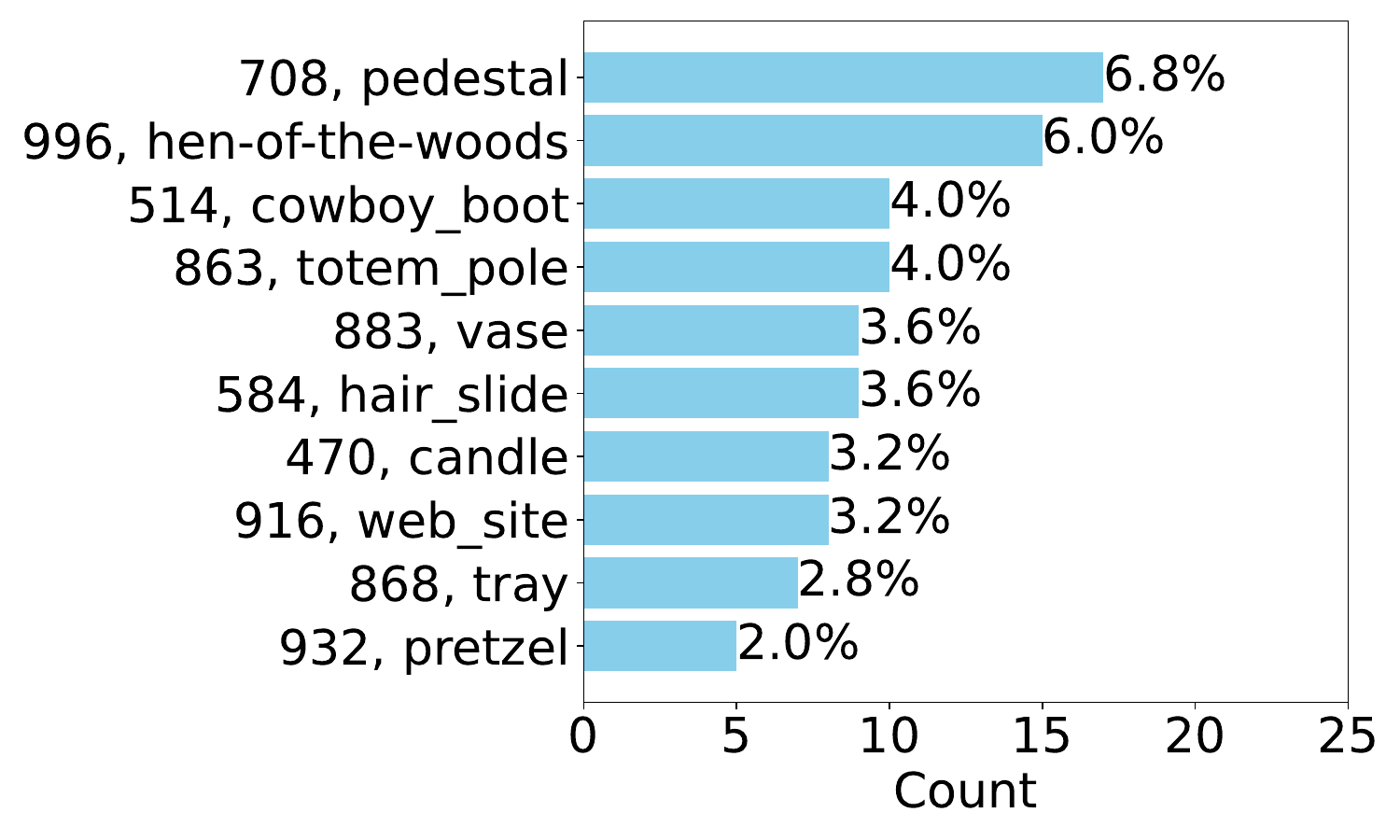}
        \label{fig:image_pos_top10}
    }
    \\ \raggedleft
    \subfloat{
        \includegraphics[width=.80\linewidth]{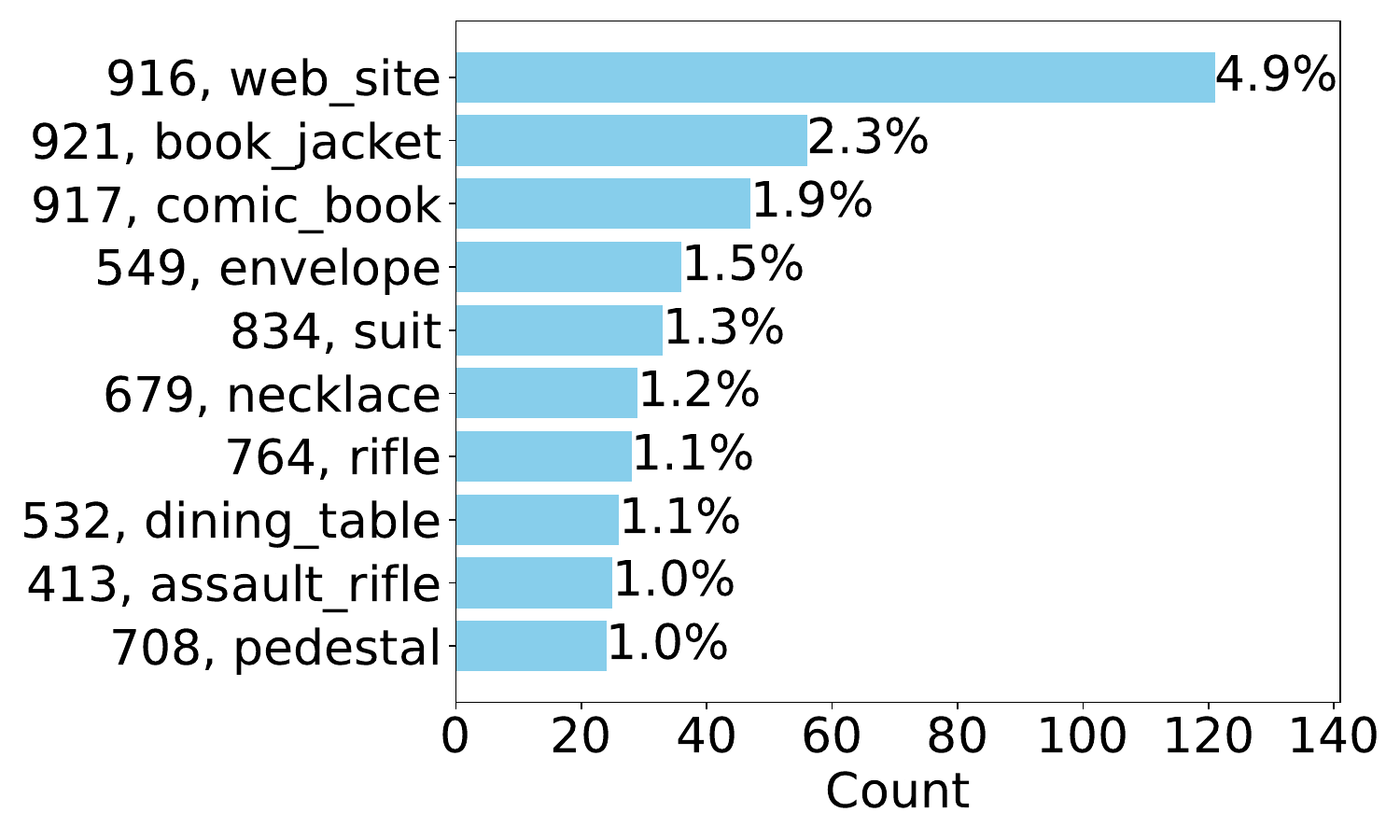}
        \label{fig:image_neg_top10}
    }
    \caption{Classification result for pos.(up) \& neg.(bottom).}
    \label{fig:image_top10}
\end{figure}

\subsubsection{Classification} 
We employed ResNet-50~\cite{he2016deep}, pretrained on IMAGENET1K\_V1~\cite{krizhevsky2012imagenet}, to classify images. 
Fig.\ref{fig:image_top10} displays the top 10 classes classified by ResNet-50 for positive and negative posts' images. 
% The histogram of the Top-10 classes is visualized in Fig.\ref{fig:image_top10}. 
% One example from each class in the Top-10 is showcased in Appendix\ref{sec:image_cls_example} (Fig.\ref{fig:example_pos} and Fig.\ref{fig:example_neg}).
% 
Positives exhibit a distinct class distribution compared to negatives. The prevalent images in positives are closely associated with the shape, material, or color of ivory carving products. In contrast, the most common images for the entire dataset and negatives are related to the ``web\_site'' class, as expected, typically containing events-related or product promotion-related information. Additionally, there are some recurring classes related to books.

\begin{figure*}[ht]
    \centering
    \small
    \subfloat[``March Estate Art, Antique ... TIMED ONLINE Auction - {{URL}} {{URL}}'']{
        \includegraphics[width=.38\columnwidth]{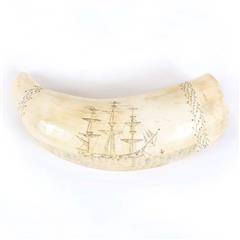}
        \label{fig:woivory}
    }
    \hspace{10pt}
    \subfloat[``... Ivory and Rhinoceros antiquities ... sold, paid and collected ... Despite the potential ban, the market is still strong ... {{URL}}'']{
        \includegraphics[width=.38\columnwidth]{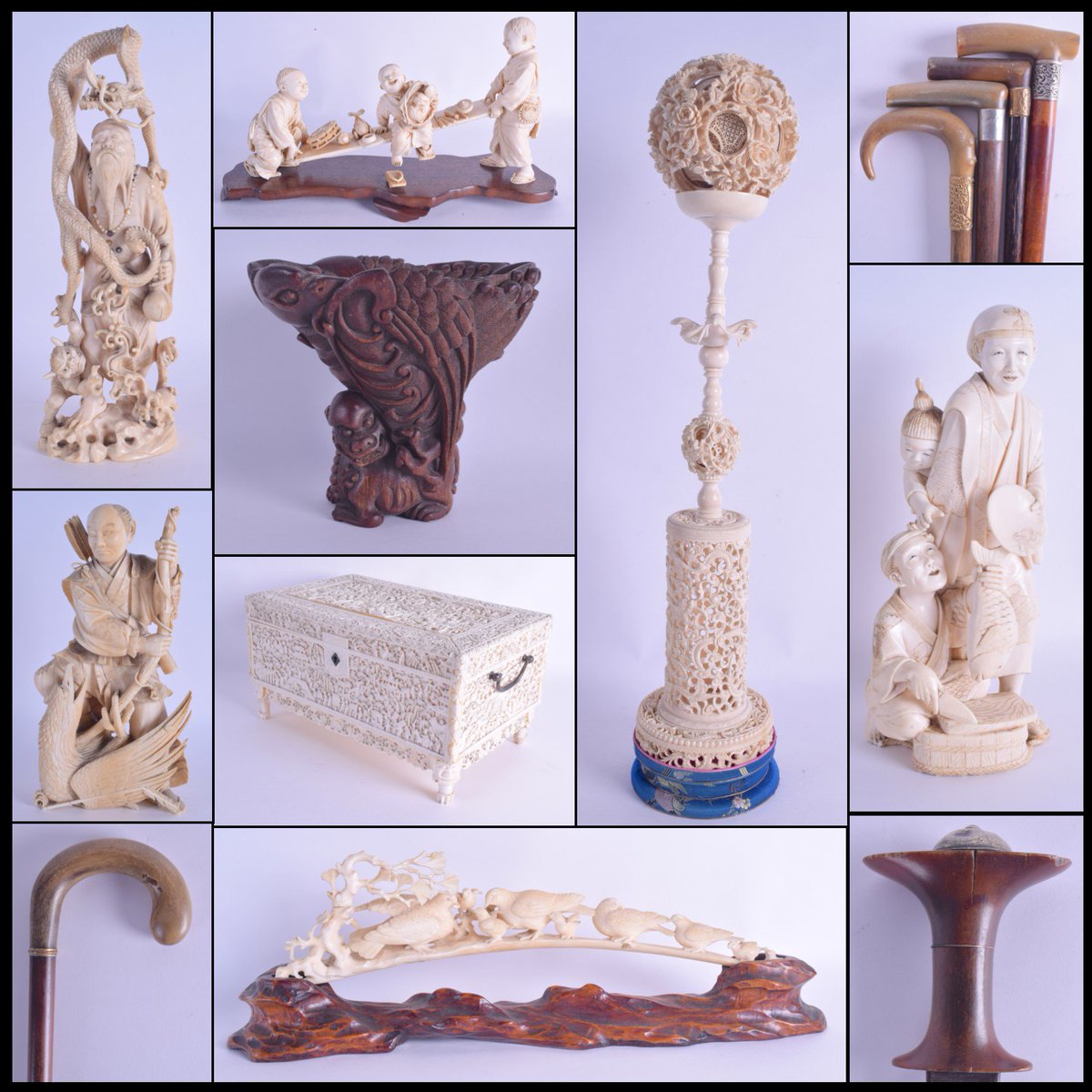}
        \label{fig:rhiino}
    }
    \hspace{10pt}
    \subfloat[``... 3 wonderfully detailed ... ancient walrus ivory. For sale on website. {{URL}}'']{
        \includegraphics[width=.38\columnwidth]{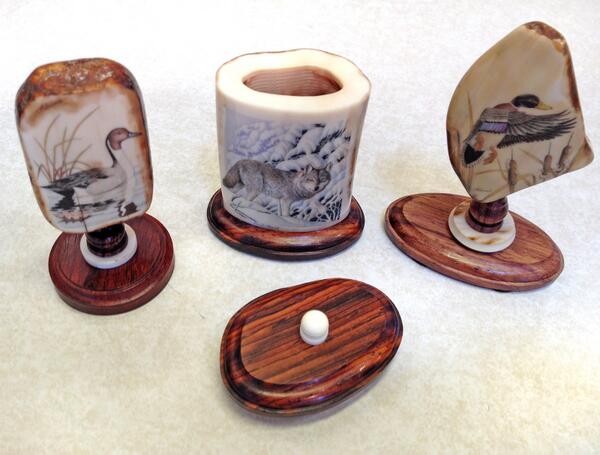}
        \label{fig:walrus}
    }
    \hspace{10pt}
    \subfloat[``Mammoth Ivory Carvings Figurine of Japanese Samurai ... precisely sculpted with original mammoth ivory. {{URL}} {{URL}}]{
        \includegraphics[width=.39\columnwidth]{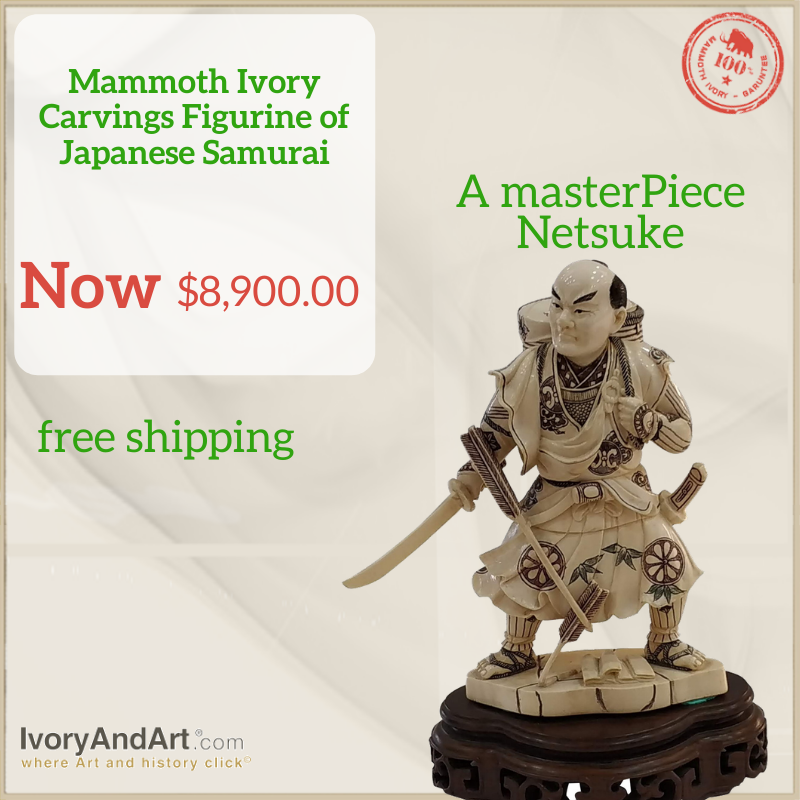}
        \label{fig:content}
    }
    \\
    \subfloat[``HB5578 will require documentation the ivory ... hoping that will save elephants in Africa??!! {{URL}}'']{
        \includegraphics[width=.39\columnwidth]{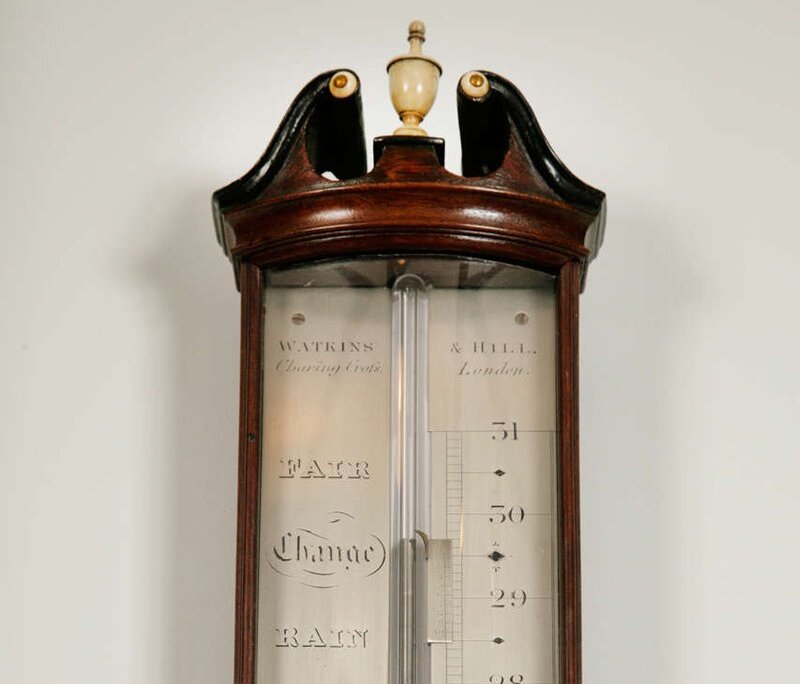}
        \label{fig:other}
    }
    \hspace{10pt}
    \subfloat[``... This Antiqued French Candelabra is ... in a beautiful patina of grey, gold and ivory. {{URL}} {{URL}}'']{
        \includegraphics[width=.38\columnwidth]{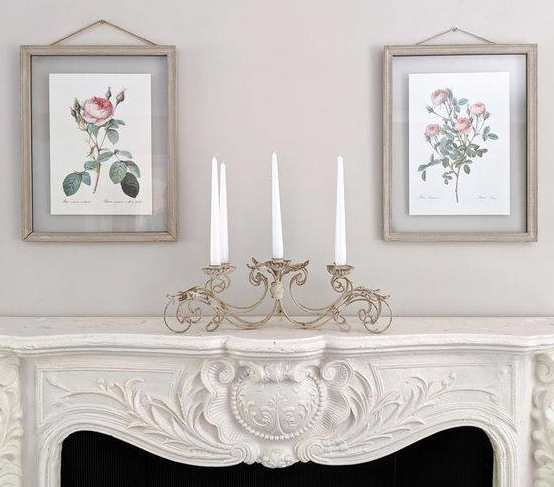}
        \label{fig:color}
    }
    \hspace{10pt}
    \subfloat[``Sumptuous Ivory Silk Quilt Bedspread {{URL}} {{URL}}'']{
        \includegraphics[width=.38\columnwidth]{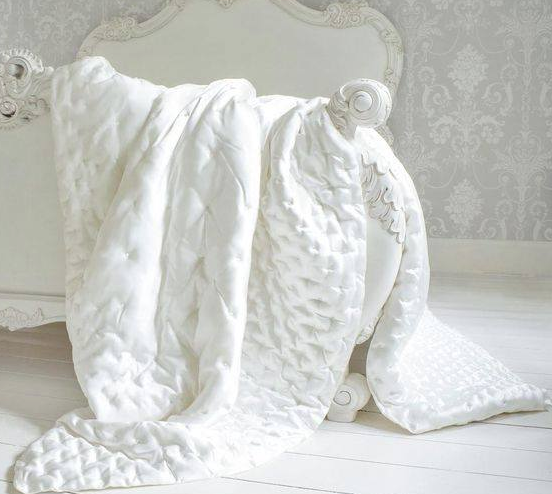}
        \label{fig:color2}
    }
    \hspace{10pt}
    \subfloat[``Special thank you to ... followers from Ivory Coast ... {{URL}} {{URL}}]{
        \includegraphics[width=.38\columnwidth]{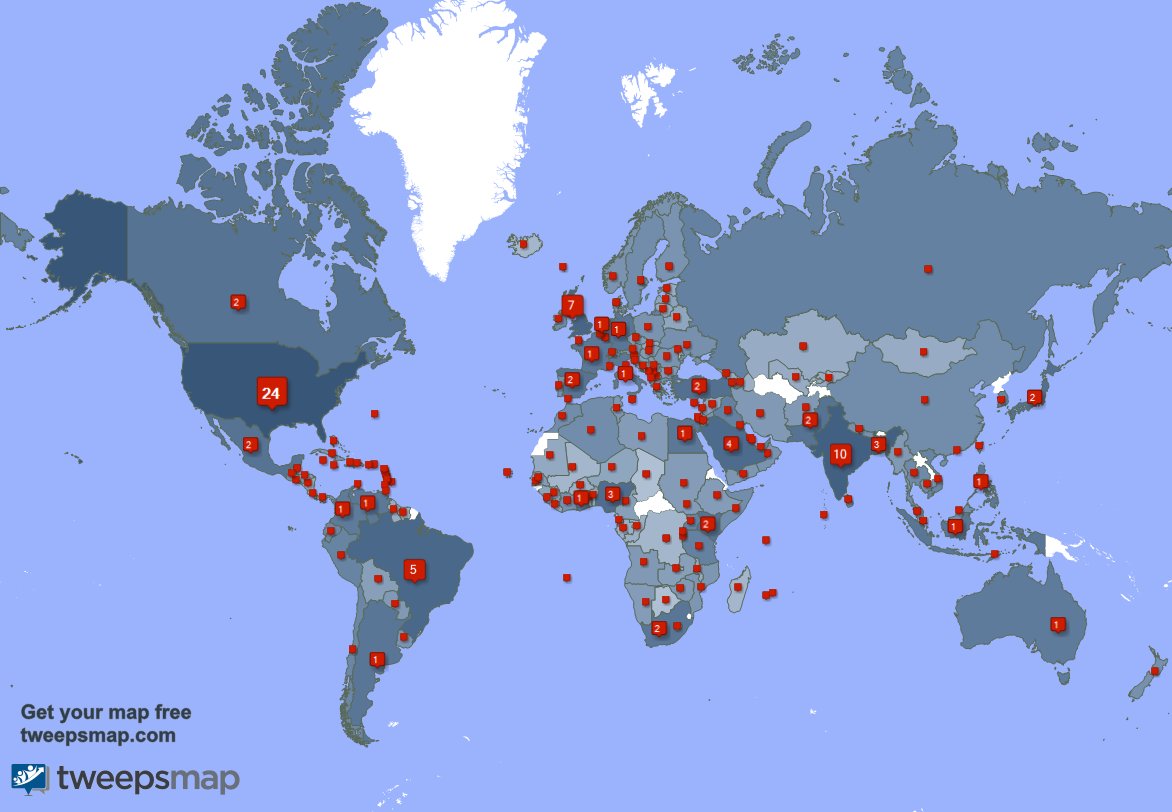}
        \label{fig:region}
    }
    \caption{WLT posts (a. -- d.) and normal posts (e. -- h.) for case study. }
    \label{fig:post}
    % \vspace{-3mm}
\end{figure*}

\section{Case Study}
\label{sec:case_study}
In this section, we present non-trivial examples to underscore the complexity of the wildlife product trading problem. Fig.~\ref{fig:post} showcases various instances, with the first row illustrating different Wildlife Product Trading (WLT) posts and the second row depicting typical normal posts.

\subsection{Positive Cases}
\label{sec:pos}
In addition to the WLT post shown at the beginning of the paper in Fig.~\ref{fig:wivory}, we explore further examples down below.

\noindent\underline{\textit{WLT without ``ivory'' in text:}}
Fig.\ref{fig:woivory} exemplifies that a WLT post may not necessarily contain the term ``ivory''. This challenges naive word filtering methods and models biased toward the term ``ivory'' in the positive class, potentially leading to false negatives. Traders may deliberately avoid explicit terms, opting for alternative ``code words'' to evade detection\cite{alfino2020code}.

\noindent\underline{\textit{WLT but not elephant ivory:}}
Ivory is not exclusive to elephants; it can originate from various animals. Fig.\ref{fig:rhiino} and Fig.\ref{fig:walrus} showcase examples of WLT posts involving products from animals such as walruses and rhinos. This highlights the involvement of WLT product sellers in multiple illegal sales activities.

\noindent\underline{\textit{WLT with sales information embedded in images:}}
Fig.~\ref{fig:content} demonstrates cases where extensive information is embedded in images, and the post class can be inferred based on images alone. This information often includes product type, price, and selling links, providing essential elements for identifying WLT-related posts.

\subsection{Negative Cases}
\label{sec:neg}
Our dataset includes many non-trivial hard-negative cases, presenting challenges in identification:

\noindent\underline{\textit{Displaying ivory products but not for trades:}}
Fig.~\ref{fig:other} shows an account engaging in a discussion on whether ivory products should be banned. This post serves a different purpose than trading and is not labeled as a WLT post.

\noindent\underline{\textit{Ivory as a color:}}
Ivory can refer to a color close to white, causing confusion for word filtering methods. Fig.\ref{fig:color} and Fig.\ref{fig:color2} exemplify instances where the term ``ivory'' post challenge in distinguishing contexts.

\noindent\underline{\textit{Special Phrases:}}
Fig.~\ref{fig:region} includes the term ``Ivory Coast'', the official name of the Republic of Côte d'Ivoire. Special phrases like this can lead to false positives for models biased on ``ivory'' without capturing the post's actual context.

\noindent\underline{\textit{Other Cases:}}
During our research, we encountered posts related to products like ``rosewood''(the ivory of forest), often associated with concerns and protection. While these are not considered positive cases in our research unless occurring simultaneously with ivory products, they highlight additional challenges in classifying wildlife product-related posts.

\section{Limitation}
\label{sec:limitation}
Our research is constrained by several limitations. Our collected dataset mainly focuses on a specific type of wildlife product: ivory-related. While there may be some differences between other types of wildlife products, our data collection, filtering, labeling, and classification modeling methods are primarily adaptable to recognizing other wildlife products. Our data collection method is network-propagation-based. Thus, the characteristics of the other isolated network clusters may not be best captured. However, given any other clusters' seed posts, we can extend the scope and establish a similar dataset given our designated approach. Lastly, it is hard to identify the actual trading record, as the real trade is accomplished on the redirected e-commerce websites or offline sites. We thus define our identification scope as the advertising for potential or finished sales and the malicious intention of selling/buying these products.

\section{Conclusion}
\label{sec:conclusion}
% We investigate wildlife product trading behaviors on online social networks, developing novel methods for data collection, filtering, and labeling. Our work yields the first dataset on ivory-product trades in the online social network. We thoroughly analyze the proposed dataset and benchmark the machine learning results based on the data. Our approach is further expandable in scale and adaptable to other wildlife product trade identification tasks with minimum extra effort. We are sharing the collected dataset to encourage future research in fighting against wildlife trafficking online behaviors. For future work, we plan to collaborate with other domain experts and help build more scalable datasets, with further potential methods using information retrieval technologies and deep learning methods. We may also design better-performing algorithms based on the proposed dataset. Notably, we may explore more options in encoding visual and text information. Although this research is laser-focused on wildlife product trading behaviors in online social networks, we note that it possess the potential in applying to other criminal marketplaces detection problems and leave these endeavours for future work.

Our investigation delves into wildlife product trading behaviors on online social networks, pioneering novel methodologies for data collection, filtering, and labeling. This endeavor culminates in the creation of the inaugural dataset focusing on ivory-product trades within the online social network sphere. Through meticulous analysis of the proposed dataset, we rigorously benchmark machine learning results based on the collected data. Notably, our approach exhibits scalability and adaptability to other wildlife product trade identification tasks with minimal additional effort.
%
% We are committed to advancing research in combating wildlife trafficking online behaviors by openly sharing the collected dataset. 
For future endeavors, we envision collaborations with domain experts to develop more expansive datasets, leveraging information retrieval technologies and deep learning methodologies. Additionally, we aim to design enhanced algorithms based on the insights gleaned from the proposed dataset.
While our research concentrates on WLT behaviors in online social networks, we note its potential in adapting to other criminal marketplaces. We leave these avenues for exploration in future research endeavors.

% Use \bibliography{yourbibfile} instead or the References section will not appear in your paper
% \clearpage
\bibliography{aaai24}

% \clearpage
\appendix

\section{Paper Checklist}
\begin{enumerate}
\small
\item For most authors...
\begin{enumerate}
    \item  Would answering this research question advance science without violating social contracts, such as violating privacy norms, perpetuating unfair profiling, exacerbating the socio-economic divide, or implying disrespect to societies or cultures?
    \answerYes{Yes. Throughout the research process, we respect user privacy and collected public data from online social networks. We tried to fight against wildlife trafficking, which is widely prohibited across majority of regions.}
  \item Do your main claims in the abstract and introduction accurately reflect the paper's contributions and scope?
    \answerYes{Yes.}
   \item Do you clarify how the proposed methodological approach is appropriate for the claims made? 
    \answerYes{Yes.}
   \item Do you clarify what are possible artifacts in the data used, given population-specific distributions?
    \answerYes{Yes.}
  \item Did you describe the limitations of your work?
    \answerYes{Yes.}
  \item Did you discuss any potential negative societal impacts of your work?
    \answerNo{No. We did not find any potential negative societal impacts of this work.}
      \item Did you discuss any potential misuse of your work?
    \answerNo{No. We did not find any potential misuse of our work.}
    \item Did you describe steps taken to prevent or mitigate potential negative outcomes of the research, such as data and model documentation, data anonymization, responsible release, access control, and the reproducibility of findings?
    \answerYes{Yes.}
  \item Have you read the ethics review guidelines and ensured that your paper conforms to them?
    \answerYes{Yes.}
\end{enumerate}

\item Additionally, if your study involves hypotheses testing...
\begin{enumerate}
  \item Did you clearly state the assumptions underlying all theoretical results?
    \answerNA{NA.}
  \item Have you provided justifications for all theoretical results?
    \answerNA{NA.}
  \item Did you discuss competing hypotheses or theories that might challenge or complement your theoretical results?
    \answerNA{NA.}
  \item Have you considered alternative mechanisms or explanations that might account for the same outcomes observed in your study?
    \answerNA{NA.}
  \item Did you address potential biases or limitations in your theoretical framework?
    \answerNA{NA.}
  \item Have you related your theoretical results to the existing literature in social science?
    \answerNA{NA.}
  \item Did you discuss the implications of your theoretical results for policy, practice, or further research in the social science domain?
    \answerNA{NA.}
\end{enumerate}

\item Additionally, if you are including theoretical proofs...
\begin{enumerate}
  \item Did you state the full set of assumptions of all theoretical results?
    \answerNA{NA.}
	\item Did you include complete proofs of all theoretical results?
    \answerNA{NA.}
\end{enumerate}

\item Additionally, if you ran machine learning experiments...
\begin{enumerate}
  \item Did you include the code, data, and instructions needed to reproduce the main experimental results (either in the supplemental material or as a URL)?
    \answerYes{Yes.}
  \item Did you specify all the training details (e.g., data splits, hyperparameters, how they were chosen)?
    \answerYes{Yes.}
     \item Did you report error bars (e.g., with respect to the random seed after running experiments multiple times)?
    \answerYes{Yes.}
	\item Did you include the total amount of compute and the type of resources used (e.g., type of GPUs, internal cluster, or cloud provider)?
    \answerYes{Yes.}
     \item Do you justify how the proposed evaluation is sufficient and appropriate to the claims made? 
    \answerYes{Yes.}
     \item Do you discuss what is ``the cost`` of misclassification and fault (in)tolerance?
    \answerYes{Yes. We provided several metrics measuring and reflecting the mis-classify rates.}
  
\end{enumerate}

\item Additionally, if you are using existing assets (e.g., code, data, models) or curating/releasing new assets...
\begin{enumerate}
  \item If your work uses existing assets, did you cite the creators?
    \answerYes{Yes. We tested several baselines as existing models. We leveraged pretrained image and text encoders in our framework. All of the above-mentioned models are cited.}
  \item Did you mention the license of the assets?
    \answerNA{NA.}
  \item Did you include any new assets in the supplemental material or as a URL?
    \answerNo{No.}
  \item Did you discuss whether and how consent was obtained from people whose data you're using/curating?
    \answerYes{Yes. All data we retrived are from online social networks, with public API access for research purpose. All models we leveraged are allowed for research purpose.}
  \item Did you discuss whether the data you are using/curating contains personally identifiable information or offensive content?
    \answerYes{Yes. During our research, we masked personally identifiable information to respect privacy. We did not find strong evidence for offensive content, as is partially shown in our sentiment analysis section.}
\item If you are curating or releasing new datasets, did you discuss how you intend to make your datasets FAIR)?
\answerYes{Yes.}
\item If you are curating or releasing new datasets, did you create a Datasheet for the Dataset (see \citet{gebru2021datasheets})? 
\answerYes{Yes. More details are shown in the data repository.}
\end{enumerate}

\item Additionally, if you used crowdsourcing or conducted research with human subjects...
\begin{enumerate}
  \item Did you include the full text of instructions given to participants and screenshots?
    \answerNA{NA}
  \item Did you describe any potential participant risks, with mentions of Institutional Review Board (IRB) approvals?
    \answerNA{NA}
  \item Did you include the estimated hourly wage paid to participants and the total amount spent on participant compensation?
    \answerNA{NA}
   \item Did you discuss how data is stored, shared, and deidentified?
   \answerNA{NA}
\end{enumerate}

\end{enumerate}
% \clearpage
\begin{figure*}[th]
  \centering
  \includegraphics[width=\textwidth]{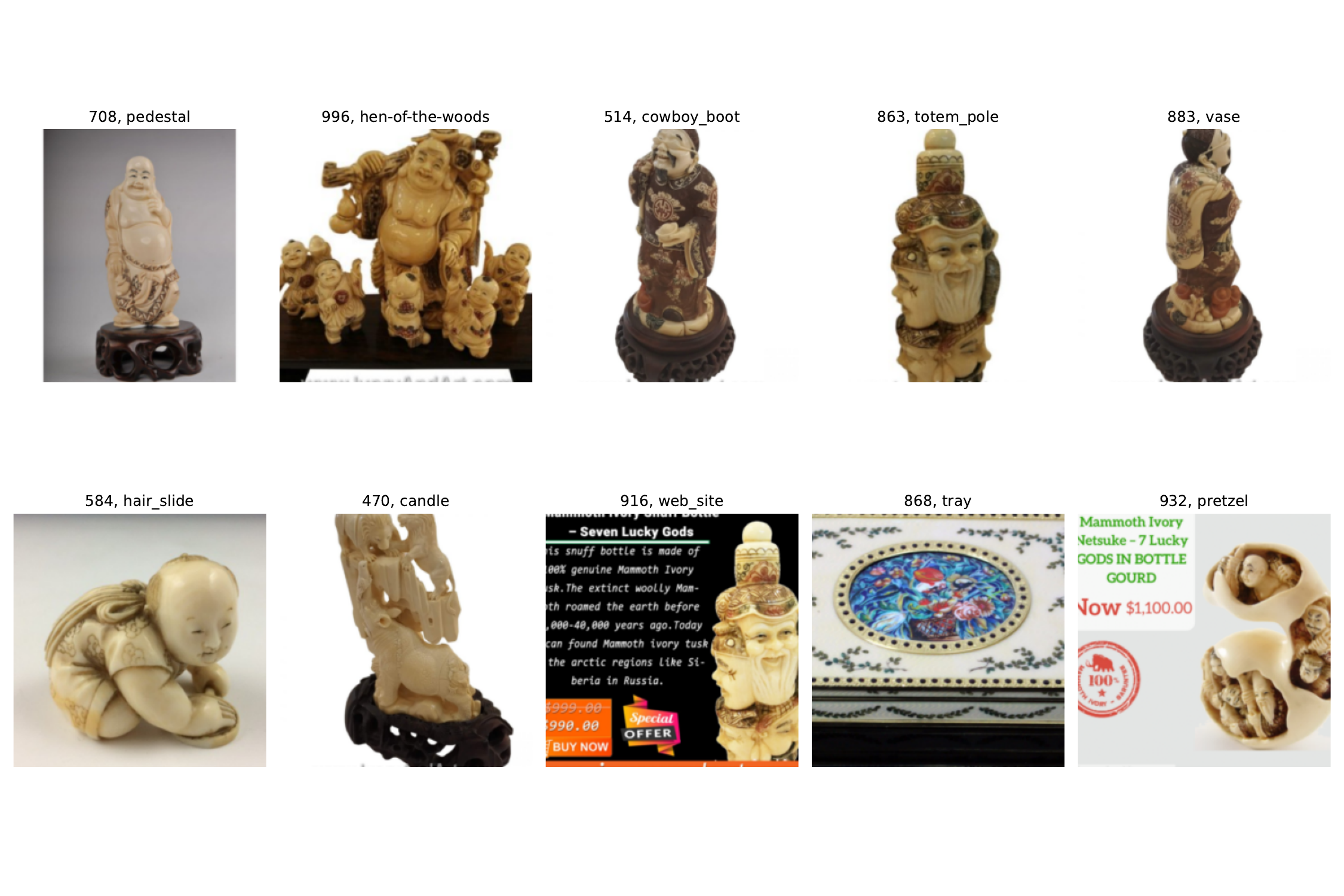}
  \caption{Top-10 image class examples in the positive/WLT class.}
    \label{fig:example_pos}
\end{figure*}

\begin{figure*}[tbh]
  \centering
  \includegraphics[width=\textwidth]{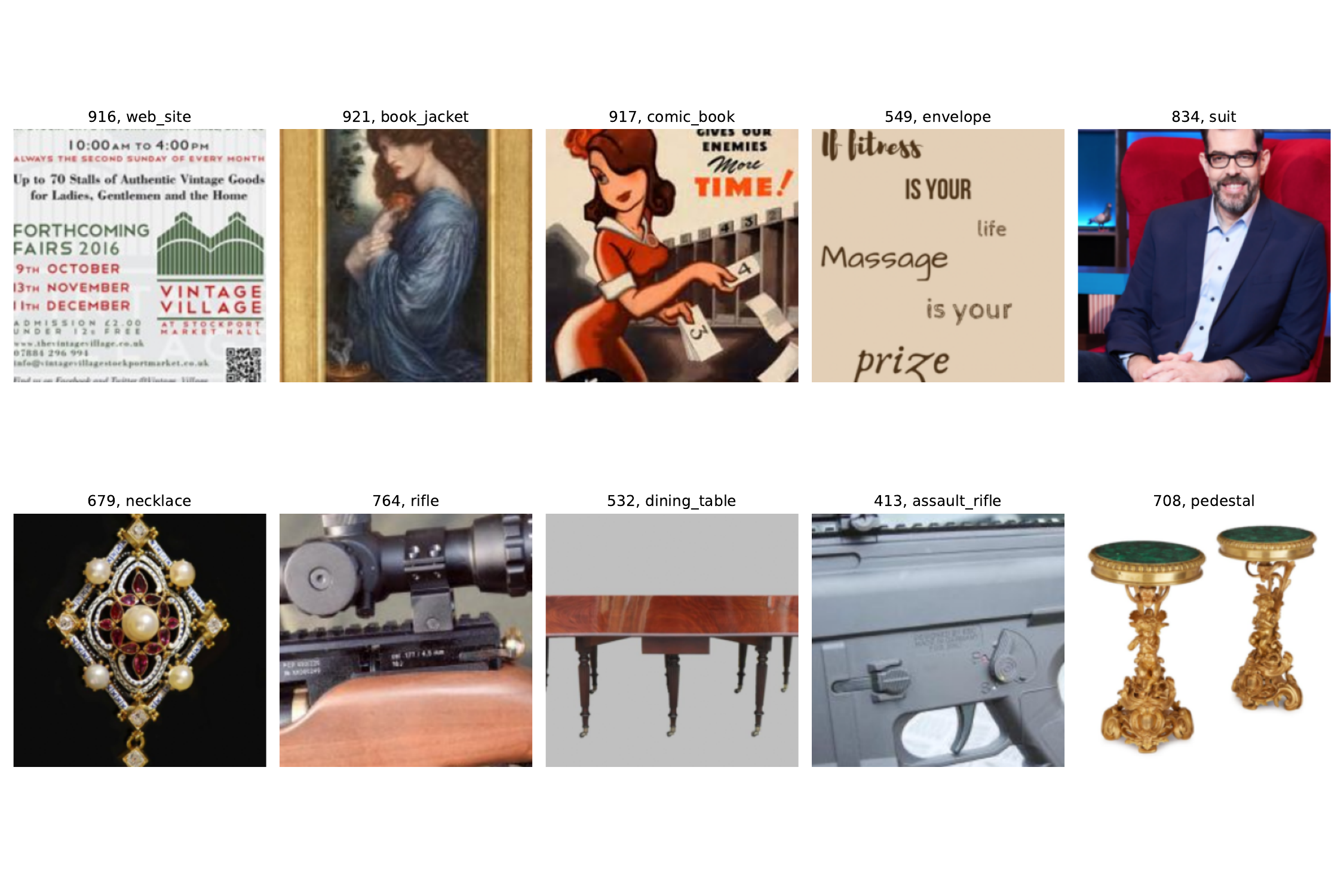}
  \caption{Top-10 image class examples in negative/normal class.}
      \label{fig:example_neg}

\end{figure*}

\section{Image Classification Examples}
\label{sec:image_cls_example}

\begin{table}[ht]
    
	\centering
    \small
    \scalebox{1.}{
	\begin{tabular}{l|l|l|l} \hline
              \textbf{Positive Class} &    \textbf{Count} & \textbf{Negative Class} &    \textbf{Count} \\ \hline
              pedestal & 17 & web\_site  &121 \\ \hline
              hen\-of\-the\-woods  & 15 &book\_jacket   &56 \\ \hline
              cowboy\_boot  & 10  & comic\_book  & 47\\ \hline
              totem\_pole  & 10  & envelope  &36 \\ \hline
              vase  & 9  & suit  & 33\\ \hline
              hair\_slide  &  9 & necklace  &29 \\ \hline
              candle & 8  &rifle   &28 \\ \hline
              web\_site  & 8  &dining\_table   &26 \\ \hline
              tray  & 7  &  assault\_rifle & 25\\ \hline
              pretzel  & 5   & pedestal  &24 \\ \hline
	\end{tabular}}
\caption{Top-K Image Classification Result.}
\label{tab:img_class}
\end{table}

\begin{figure*}[ht]
    \centering
    \small
    \subfloat[WLT Users]{
        \includegraphics[width=.32\linewidth]{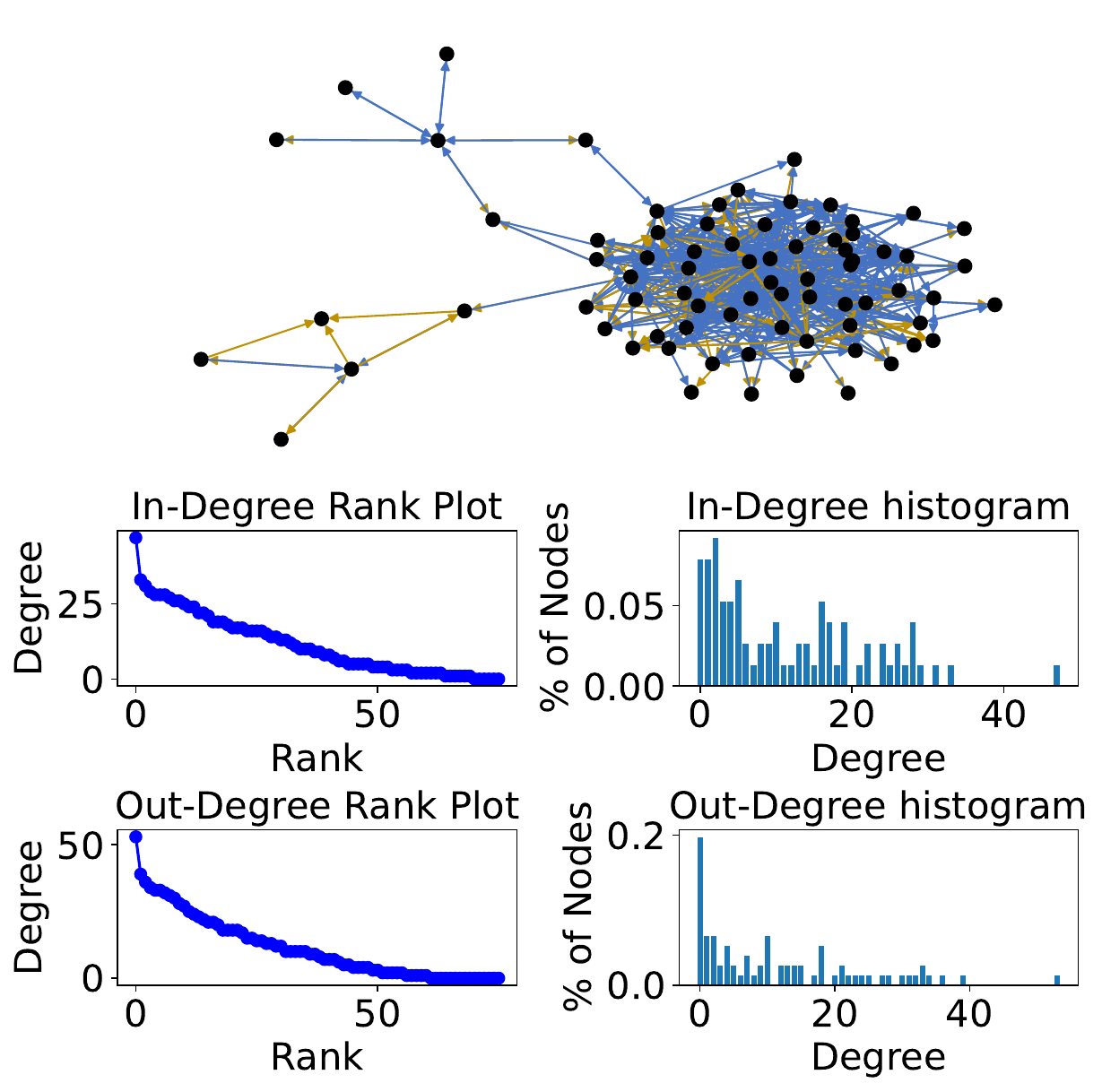}
        \label{fig:wlt_network}
    }
    \subfloat[Normal Users]{
        \includegraphics[width=.32\linewidth]{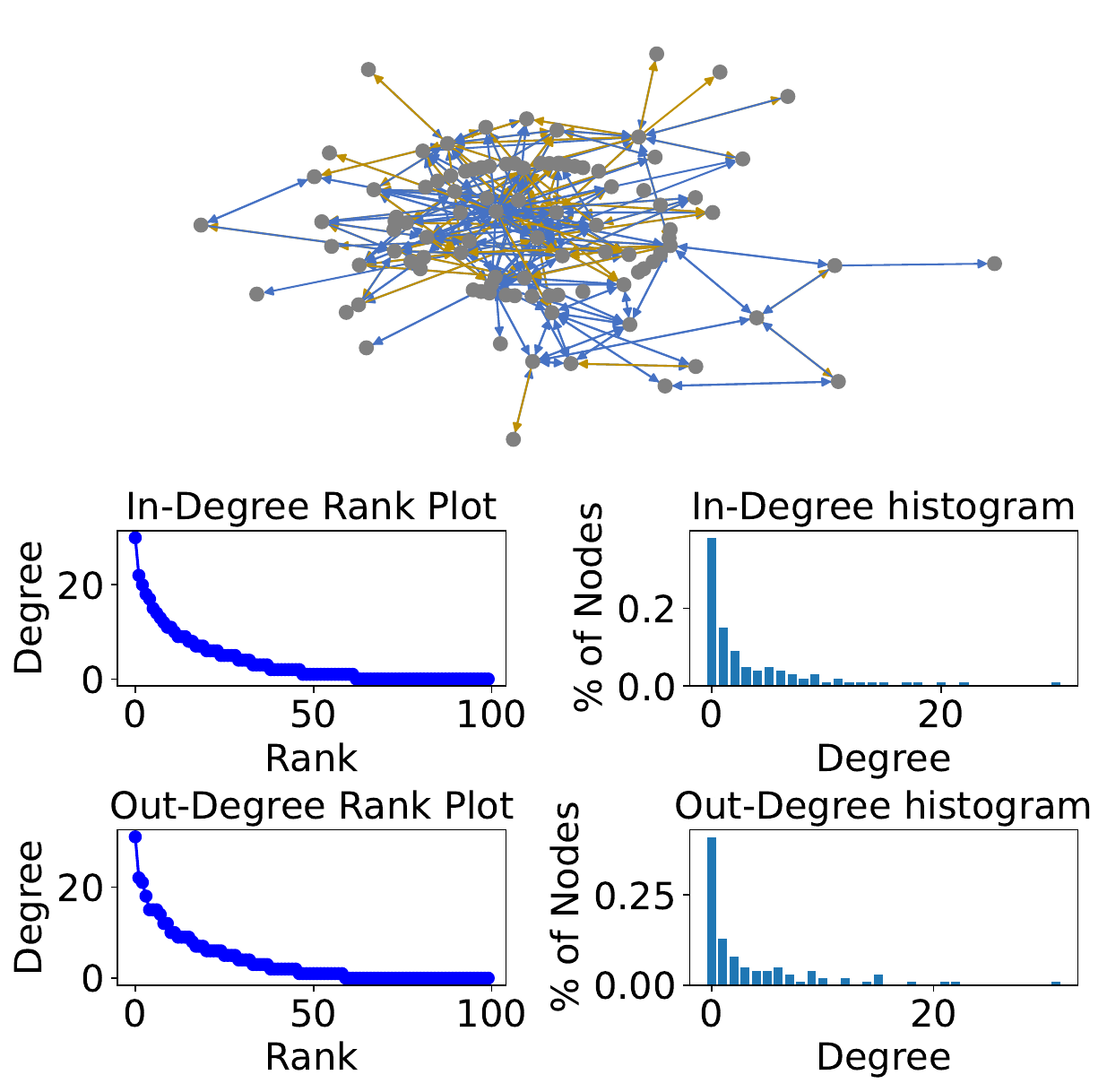}
        \label{fig:normal_network}
    }
    \subfloat[Network between two classes]{
        \includegraphics[width=.32\linewidth]{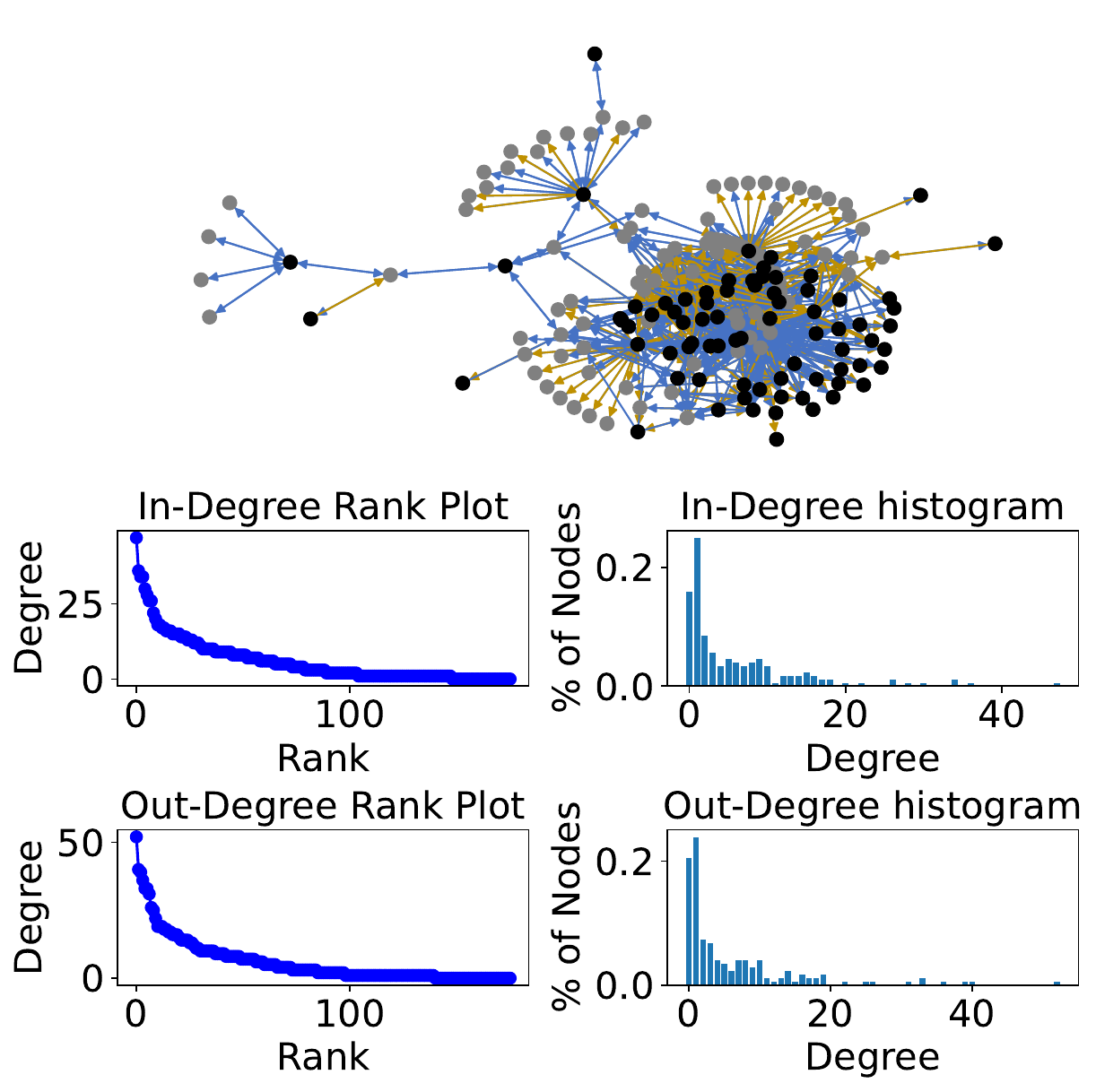}
        \label{fig:intra_network}
    }
    \caption{Connections and distributions of users.}
    \label{fig:user_analysis}
\end{figure*}

\section{Posting Users}
\label{sec:data_user}
We provide some straightforward analysis on the posting users in Fig.~\ref{fig:user_analysis}. We define the ``WLT users'' as those had at least one WLT post, and the others as ``normal users''. Note that WLT users can also post normal posts. We plot the inter-connections among WLT users in Fig.~\ref{fig:wlt_network}, the inter-connections among normal users in Fig.~\ref{fig:normal_network}, and the intra-connections between WLT users and normal users in Fig.~\ref{fig:intra_network}. The following and follower connections are shown in two different edge colors. To specify the differences, we plot the in-degree and out-degree distributions separately. Note that we downsampled the normal users to 100 such that they have the similar scale in \# users as WLT users.  We observe that WLT users tend to have higher in-degree and out-degree extreme cases (the left-most parts in rank plots). However, in a more general scope, the WLT users tend to have less connections than the normal users.

We are aware of other user definitions (e.g., generalize at least one WLT-post to at least $k$ WLT-posts) or other user analysis perspectives (e.g., username, user profile description, user verified status, user profile image analysis, etc.). We focused on the network analysis and leave the other analysis for future work. Although the analysis opens a new discussion opportunity around user-level wildlife product trading identification task, this paper mainly focused on post-level identification problem.

\end{document}